\documentclass[preprintnumbers,superscriptaddress,nofootinbib,aps,prd,floatfix]{revtex4}

\usepackage{wrapfig}
\usepackage{hyperref,color}
\usepackage{amsmath,slashed}
\usepackage{graphicx}

\newcommand{\be}{\begin{eqnarray*}}
\newcommand{\ee}{\end{eqnarray*}}

\newcommand{\bee}{\begin{eqnarray}}
\newcommand{\eee}{\end{eqnarray}}
\newcommand{\beeq}{\begin{equation}}
\newcommand{\eeeq}{\end{equation}}

\newcommand{\mg}{\texttt{MadGraph\_aMCatNLO}}
\newcommand{\hw}{\texttt{Herwig~7}}
\newcommand{\fj}{\texttt{FastJet}}
\newcommand{\tmva}{\texttt{TMVA}}
\newcommand{\madmax}{\texttt{MadMax}}

\begin{document}

\title{Same-sign $W$ pair production in composite Higgs models}

\begin{abstract}
Non-minimal composite Higgs scenarios can contain exotic Higgs states which, if getting
observed at the Large Hadron Collider, will help to constrain the underlying UV structure
of the strong dynamics. Doubly charged Higgs bosons are well-motivated scalar degrees of freedom in this context.
Their phenomenology in typical composite scenarios can differ from well-established Higgs triplet
extensions of the SM. Related search strategies are not necessarily adapted to such a scenario as
a consequence. In this paper we discuss the sensitivity reach to doubly charged Higgs
bosons with decays into pairs of same-sign $W$ bosons. While production cross sections are small, we show
that significant constraints on $H^{\pm\pm}\to W^{\pm}W^{\pm}$ can be obtained, providing a new opportunity to constrain the potential composite structure of the TeV scale up to $m_{H^{\pm\pm}}\simeq 800$~GeV.
\end{abstract}

\author{Christoph Englert}

\affiliation{SUPA, School of Physics and Astronomy, University of
  Glasgow, Glasgow G12 8QQ, UK}
\author{Peter Schichtel }
\affiliation{Institute for Particle Physics Phenomenology, Department
  of Physics, Durham University, Durham DH1 3LE, UK}
\author{Michael Spannowsky}
\affiliation{Institute for Particle Physics Phenomenology, Department
  of Physics, Durham University, Durham DH1 3LE, UK}

\preprint{IPPP/16/92}
\preprint{DCPT/16/184}
\preprint{MCnet-16-39}

\maketitle

\section{Introduction}
\label{sec:intro}

After the Higgs boson discovery, followed by the lack of any conclusive
hints for physics beyond the Standard Model (BSM), the hierarchy
problem remains one of the most pressing nuisances that our
understanding of the TeV scale faces. Although TeV scale naturalness
is not a technical problem, it is surprising that the
success story of perturbative quantum field theory seems to be challenged
by a state whose appearance is directly linked to the TeV scale
itself. Many BSM scenarios have been devised over the past decades to
explain the TeV scale as a natural phenomenon either through
perturbative cancellations guaranteed by approximate (super)symmetry or
through dimensional transmutation effects. The latter 
adapt ideas of QCD confinement to the TeV scale with phenomenologically 
necessary model-building adjustments.

It is fair to say that there is a much more detailed understanding of
supersymmetric extensions of the SM than there is for strongly
interacting theories of the TeV scale. This is mostly due to the fact
that perturbative methods are bound to break down for strongly
interacting theories and phenomenological applications necessarily
need to revert to chiral perturbation theory ($\chi$PT)
techniques~\cite{Gasser:1984gg,Leutwyler:1991mz,Leutwyler:1993iq}. Although $\chi$PT is an extremely successful concept (for
instance, it serves to explain the pion mass splitting~\cite{Contino:2010rs} which
acts as a blueprint for composite Higgs scenarios~\cite{Contino:2003ve,Contino:2006nn,Agashe:2004rs,Contino:2010rs,Gillioz:2012se,Giudice:2007fh}), it is
unclear whether UV completions of a particular low energy theory do
indeed exist. This constitutes a long-standing problem in the classification of composite Higgs scenarios, which has motivated investigations using
both dualities~\cite{Faedo:2013ota,Athenodorou:2016ndx} and lattice computations~\cite{Golterman:2015zwa,DeGrand:2015zxa,DeGrand:2016htl}.

While UV completions of the minimal composite Higgs (MCHM) scenarios have proven
difficult to construct, there are non-minimal models
with known UV completions~\cite{Ferretti:2013kya,Barnard:2013zea,Ferretti:2014qta}. Typically,
these non-minimal extensions predict a range of scalar~\cite{Cacciapaglia:2015eqa,Belyaev:2016ftv}, fermionic~\cite{Agashe:2006at} as
well as potentially vectorial exotics. Among these, scalar exotics such as
a doubly charged Higgs boson are tell-tale stories of the
compositeness nature of the TeV scale~\cite{Mrazek:2011iu}. 

Such states are searched for by the ATLAS and CMS experiments, either in
leptonic decays, motivated from generic Higgs $SU(2)_L$
triplet extensions~\cite{Akeroyd:2009hb,Akeroyd:2010ip}, or same sign $W$ boson final states. The latter will be induced if the triplet plays a role in electroweak symmetry
breaking~\cite{Georgi:1985nv,Chanowitz:1985ug,Gunion:1989ci,Englert:2013wga,Bambhaniya:2015wna,Logan:2015xpa,kang:2014jia,Grinstein:2013fia}.

In some composite Higgs scenarios there are doubly charged Higgs
bosons that fall into the middle of these analysis strategies: they
play no role in the vacuum misalignment which triggers electroweak
symmetry breaking and their coupling to leptons is either absent or
suppressed. It is the purpose of this work to close this gap and
provide sensitivity estimates for signatures that are motivated from these
non-minimal composite Higgs scenarios in the TeV regime.

This paper is organised as follows. To make this work self-contained,
we quickly review aspects of doubly charged Higgs bosons that arise in
composite Higgs scenarios in Sec.~\ref{sec:model}. In
Secs.~\ref{sec:events} and \ref{sec:analysis}, we give details of our
event simulation and analysis strategy before we draw conclusions in
Sec.~\ref{sec:conc}.


\section{Model and Motivation}
\label{sec:model}
Recently, a potential UV completion of a non-minimal composite Higgs
model was introduced in~\cite{Ferretti:2014qta} (see also~\cite{Ferretti:2016upr}). We will use this 
scenario as our main motivation for a search $H^{\pm\pm}\to W^\pm W^\pm$. The model of Ref.~\cite{Ferretti:2014qta} is based on the symmetry group
\begin{equation}
	\underbrace{SU(4)}_{H_c}\times \underbrace{SU(5)\times SU(3) \times SU(3)' \times U(1)_X\times U(1)'}_{G_F}\,,
\end{equation}
with fermions transforming as $\psi \in \bf{6}, \chi \in 4, \tilde \chi \in \bar 4$ under the ``hyper colour'' gauge group $H_c=SU(4)$. Strong $SU(4)$ dynamics cause the breakdown of the global symmetries  $G_F$
\begin{equation}
\label{eq:su5}
SU(5)\to SO(5)\,,
\end{equation}
and
\begin{equation}
\label{eq:su32}
SU(3)\times SU(3)' \to SU(3),
\end{equation}
as well as the breaking of $U(1)'$. The author of \cite{Ferretti:2014qta} argues that Eq.~\eqref{eq:su5} occurs at a higher scale than Eq.~\eqref{eq:su32}; the low energy effective theory can then be parametrised by the coset
\begin{equation}
	\label{eq:symmbreak}
	G_F/H_F =	
	  {SU(5)\over SO(5)} \times {SU(3) \times SU(3)'\over SU(3)} \times U(1)'\,.
\end{equation}
The unbroken global symmetry group $H_F$ contains the subgroup 
\begin{equation}
\label{eq:pregauge}
H_F \supset SU(3)_c\times SU(2)_L\times SU(2)_R\times  U(1)_X\,,
\end{equation}
which can be weakly gauged to arrive at the SM gauge structure.

The symmetry breaking pattern leaves a number of distinct exotics in the theory's spectrum (for instance there is a ``hypergluon''~\cite{Kilic:2009mi,Kilic:2010et,Schumann:2011ji} and an inert singlet). Our analysis targets the enlarged Higgs spectrum compared to MCHM4~\cite{Agashe:2004rs} or MCHM5~\cite{Contino:2006qr}. The Nambu Goldstone bosons that arise from $SU(5)\to SO(5)$ transform under $SU(2)_L\times U(1)_Y$ as 
\begin{equation}
	{\bf{1}}_0 + {\bf{2}}_{\pm 1/2} + {\bf{3}}_0 + {\bf{3}}_{\pm 1} \,,
\end{equation}
and we can interpret the ${\bf{2}}_{\pm 1/2}$ multiplet as the SM Higgs field. Weakly gauging the electroweak group as part of Eq.~\eqref{eq:pregauge}, together with the presence of a heavy top quark, induces a Coleman-Weinberg potential \cite{Coleman:1973jx} for this multiplet, which triggers electroweak symmetry breaking as the vacuum becomes dynamically misaligned with respect to the $SU(2)_L\times U(1)_Y~(Y=T^3_R+X$) preserving direction~\cite{Contino:2010rs}.

A phenomenological smoking gun of this scenario is the appearance of a ${\bf{3}}_{\pm 1}$ multiplet, which contains a doubly charged Higgs boson that, however, has no relation to the electroweak scale as vacuum misalignment proceeds entirely through ${\bf{2}}_{\pm 1/2}$ interactions. This phenomenological situation is vastly different from other Higgs triplet scenarios~\cite{Georgi:1985nv,Chanowitz:1985ug,Gunion:1989ci,Akeroyd:2005gt,Akeroyd:2010ip}: firstly, tension with the $\rho$ parameter (either in custodial~\cite{Gunion:1990dt} or non-custodial realisations~\cite{Akeroyd:2005gt}) is relaxed and related fine-tuning is absent. Secondly, since ${\bf{3}}_{0,\pm 1}$ do not participate in electroweak symmetry breaking, they will not leave an observable signature in weak boson fusion final states~\cite{Godfrey:2010qb,Cheung:2002gd,Englert:2013wga,Bambhaniya:2015wna,Logan:2015xpa,Degrande:2015xnm}, which are particularly suited to custodial Higgs triplet models. Instead their production will need to happen through pair production~\cite{Akeroyd:2005gt} entirely 
\begin{wrapfigure}[14]{r}{0.4\textwidth}	
\hfill\parbox{0.38\textwidth}{
\centering
\includegraphics[width=0.24\textwidth]{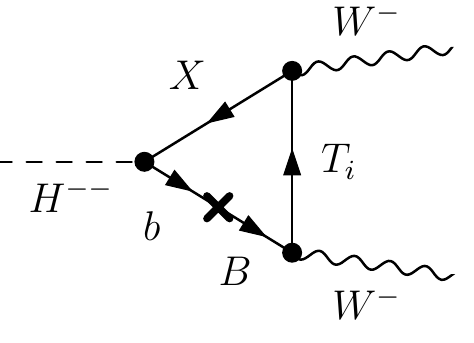}
\caption{Typical Feynman diagram contributing to the decay $H^{--}\to W^{-}W^{-}$ in the Lagrangian eigenbasis.}
\label{fig:decay}}	
\end{wrapfigure}
fixed by the quantum numbers of the weak isotriplet.

With the only distinction of Eq.~\eqref{eq:symmbreak},~the model of~\cite{Ferretti:2014qta} follows the paradigm of the MCHM scenario; massive bottom and top quarks are included through partial compositeness~\cite{Kaplan:1991dc,Contino:2006nn} by introducing three top partners $\{T_i\}$ and one bottom partner $B$ that lift the fundamental $t,b$ masses. Similar to MCHM5, this introduces a range of effective Higgs-fermion interactions. The interaction most relevant to the present work is given by \cite{Ferretti:2014qta}
\begin{equation}
	{\cal{L}}\supset -\sqrt{2}i\lambda_q \, \bar b_{L} X_R H^{--} + \text{h.c.}
\end{equation}
where $X$ denotes a top partner with charge $5/3$ that is characteristic for custodial symmetry preserving composite Higgs scenarios~\cite{Agashe:2006at}. The $X$ field together with four other top-like fields in MCHM5 forms a charged current that couples to the $W$ field. The decay of the doubly charged Higgs boson into same sign $W$s, albeit absent at tree-level, hence proceeds at loop level through including a mass insertion, as indicated in Fig.~\ref{fig:decay}. This decay is therefore directly related to the mixing angles that generate the physical $b$ quark mass\footnote{This is similar to the decay $H\to \gamma\gamma,gg$ in the SM, with the photons and gluons standing in no relation to spontaneous symmetry breaking.}. Note, that there is also the possibility of this decay to be anomaly term induced \cite{Ferretti:2016upr} with a different CP and Lorentz structure.\footnote{We are grateful to Gabriele Ferretti for pointing out this possibility.} In practice the decay amplitude can be evaluated through rotating the top quark partners to the mass-eigenbasis which results in a non-diagonal coupling structure of the $W$ in the space of top and bottom quarks. The very decay of the doubly charged Higgs into same sign $W$ therefore carries a lot of (although degenerate) information about the particular structure of mass generation through partial compositeness. 

Numerically there is only a small mixing required to lift an elementary fermion to the mass of the bottom quark, and in general one expects that the decay amplitude is parametrically small (in line with the effective field theory analysis of~\cite{Azatov:2013ura}). The difference compared to the minimal composite scenarios in the present case, however, is the potential absence of low-energy two-body final states that allow a prompt and charge-conserving $H^{\pm\pm}$ decay.\footnote{The composite dynamics that will create a dynamical mass for $H^{--}$ will also lift $H^{-}$ and mass splittings can occur inside the $T^{3}_L=1$ multiplet or between the $|Y|=0,1$ multiplets. However, unless the strong dynamics changes the picture radically (which need to be assessed on the lattice), the mass splittings should be entirely perturbative and qualitatively similar to the $\pi^\pm-\pi^0$ mass splitting. From the perspective of chiral perturbation theory cascade decays $H^{++}\to H^+ H^+$ seem unlikely. Current constraints exclude top partners in the range of 600 GeV~\cite{Matsedonskyi:2015dns}. We therefore assume $m_X>m_{H^{\pm\pm}}$ in the following.}

 While details of the doubly charged Higgs decays are certainly model-dependent and maybe even beyond perturbative control, it is clear that any statistical significant observation or exclusion of $H^{\pm\pm}\to W^\pm W^\pm$ leading to a distinctive resonant final state 
\begin{equation}
\label{eq:signature}
	  l^+\,l^+\,l^-\,l^-\,\slashed{E}_T\,,
\end{equation}
with Higgs masses in the multi-hundred GeV regime, will have strong implications for the underlying effective description of the compositeness model, with direct ramifications for its underlying UV structure (e.g. through the mixing effects in Fig.~\ref{fig:decay}).

This particular final state is not plagued by large QCD backgrounds, but signal cross sections are generically small~\cite{Kanemura:2013vxa,Kanemura:2014goa}. Including estimates for charge mis-tagging etc. is therefore important to arrive a at realistic expectation of the sensitivity to this model. Taking the pNGB character of the Higgs boson at face value, we can expect that the doubly charged Higgs boson is heavy and we will leave its mass as a free parameter $m_{H^{\pm\pm}}>125~\text{GeV}$ in our analysis. In particular we will compare the performance of various analysis approaches that make use of large discriminative power for the expected small signal vs. background ratio for large doubly charged Higgs masses. 


\section{Event Simulation}
\label{sec:events}
Focusing on the leptonic $W$ decay modes we consider
Eq.~\eqref{eq:signature} as final state signature, where the
$\slashed{E}_T$ stems from four neutrinos in the final state and
$l\in\{e,\mu\}$. As Eq.~\eqref{eq:signature} describes a very clean
channel at the LHC, the only irreducible backgrounds we consider are
$W^+W^+W^-W^-$, four lepton production in association with a $Z$, and
two leptons together with $W^+W^-$. All other backgrounds are 
reducible. We identify jet fakes, i.e a jet being mistaken as a lepton,
as the main source of such events. The fake rate we control via~\cite{ATL-PHYS-PUB-2013-004}
\begin{align}
  \text{P}\left( j\rightarrow e \right) &= 0.0048 \times \text{exp}
  \left[ -0.035 \times \frac{p_{T,j}}{\text{GeV}} \right] \, ,
\end{align}
for all jets within the reach of the electromagnetic calorimeter. In
practice, we explore jets clustered with the anti-$k_T$ algorithm, as
implemented in \fj~\cite{Cacciari:2011ma}, with a radius parameter of
$R=0.4$ and $p_T^\text{min}\ge10$~GeV. Therefore, the highest
probability to fake an electron is given by low momentum jets
$\text{P}^\text{max} \left( j\rightarrow e \right) = 3.4 \times
10^{-3}$. This gives us a direct handle to estimate whether a certain
reducible background with a given cross section is important for our
study. Two leptons plus $W^\pm$ and jet production provide candidate
events for such a background, as well as Drell Yan in association with
at least two jets, for which the jets need to fake the number of
missing leptons.  Similarly, lepton neutrino production with three
jets need to be included as well. In principle, QCD-induced jet
production contributes too, however, requiring at least one muon in
the final state, we render the latter one irrelevant, while keeping
$93\%$ of the signal.

Top quark production constitutes a further category of backgrounds. We consider $t\bar{t}$ production in association with
either $W^+W^-$ or $l^+l^-$. When there are $B$ mesons in the final state we
need to take the $b$ mis-tagging rate into account as well. We assume the
tagging rate to be $70\%$ per true
$b\,$-jet~\cite{ATL-PHYS-PUB-2015-022,Piacquadio:2008zza}; a $b$ veto
results in an efficiency factor of $0.09$. 

Due to inherent and fake
missing energy in QCD jet radiation we study the production of four
leptons as last class of backgrounds in our list of dominant backgrounds. As we do not include
a full detector simulation, we need to model the amount of missing
energy. It has been shown that
\begin{align}
  \frac{\Delta\slashed{E}_T}{\slashed{E}_T} &=
  \frac{2.92}{\slashed{E}_T} - 0.07
  \label{eq:smearing}
\end{align}
corresponds to a conservative estimate~\cite{Englert:2012wf}.
Therefore, we smear the missing energy vector as computed from Monte-Carlo truth with a gaussian distribution according to  Eq.~\eqref{eq:smearing} instead of
modeling the detector response to every single final state particle. 

A
further obstacle in describing the final state is charge miss
identification. As for the electron fakes we use a transverse momentum
dependent description~\cite{Aad:2008zzm}
\begin{align}
  P \left( \text{charge flip} \right) &= \text{min} \left( 0.2,\, 4.68
  \times 10^{-8}~ \left[\, \frac{p_T}{\text{GeV}} + 65.0 \right]^2
  \right)\,,
  \label{eq:charge_miss}
\end{align}
and we ignore the possibility of a charge flip for muons. Equation~\eqref{eq:charge_miss}
results from fitting a parabola to the data points that we have extrapolated
from Ref.~\cite{Aad:2008zzm}. We checked that the results presented in
this paper do not show a significant dependence on the exact result of this fit.

As we are interested in the main kinematic features of the relevant
channels, we simulate all processes at their respective leading order
(LO). We use \mg~\cite{Alwall:2011uj,Alwall:2014hca} to generate
matrix element level results for our model and
\hw's~\cite{Bellm:2015jjp} angular ordered parton
shower~\cite{Gieseke:2003rz}. For the simulation of the background we
employ \mg's LO amplitudes interfaced to the \hw~generator, again
using the angular ordered parton shower, with \hw~taking care of all
decays~\cite{Hamilton:2006ms,Bahr:2008pv,Gigg:2008yc}. Since QCD
details are not too important for this particular clean final state
apart from changed normalisations (see below), we do not include
hadronization and underlying event (UE) effects. Indeed, we checked
that including UE has no influence on the lepton separation for our
signal process.

In Tab.~\ref{tab:xsec_mc} we collect the simulated processes as
well as their cross section on generator level $\sigma_\text{MC}$. The
fiducial cross sections for $l^+l^+l^-l^-Z$, $t\bar{t}W^+W^-$, and
$l^\pm\nu jjj$ are indeed so small that we do not need to consider them any
further. 

For some of the processes next-to-leading order (NLO) corrections are
large. In our MC generation we therefore include the following flat
$K=\sigma^{\text{NLO}}/\sigma^{\text{LO}}$-factors: $1.3$ for all signal processes due to a Drell-Yan character~(e.g.~\cite{Altarelli:1978id}), $2.0$ for
$W^+W^+W^-W^-$~(conservatively adopted from $WWZ$ production \cite{Hankele:2007sb,Bozzi:2009ig,Campanario:2008yg}) and $l^+l^-W^+W^-$~\cite{Hankele:2007sb,Bozzi:2009ig,Campanario:2008yg} as well as $1.6$ for
$l^+l^+l^-l^-$~\cite{Campbell:2011bn,Grazzini:2015hta} and $l^+l^-W^\pm j$~\cite{Campanario:2010hp,Campanario:2009um}. For all other
processes we neglect any NLO corrections as they turn out to be negligible contributions to the background.
\begin{table}[!t]
  \renewcommand{\arraystretch}{1.5}
  \centering
  \begin{tabular}{c|cccccc}
    process                                    & ~$W^+W^+W^-W^-$~ & ~$l^+l^+l^-l^-Z$~ & ~$l^+l^+l^-l^-$~  & ~$l^+l^-W^+W^-$~ &  ~$t\bar{t}W^+W^-$~     & ~$t\bar{t}l^+l^-$~    \\
    \hline
    \hline
    $\sigma_\text{MC} [\text{ab}]$              & $2.2$           & $13$              & $21$             & $538$            & $11$                    & $1.2 \times 10^3 $  \\
    $\epsilon_\text{four leptons}$              & $0.62$          & $0.71$            & $0.37$           & $0.52$           & $0.32 \times 10^{-2}$    & $0.30 \times 10^{-2}$ \\
    $\epsilon_{++--}$                           & $0.62$          & $0.70$            & $0.37$           & $0.52$           & $0.31 \times 10^{-2}$    & $0.29 \times 10^{-2}$ \\
    $\epsilon_\text{fiducial}$                  & $0.56$          & $0.69$            & $0.36$           & $0.48$           & $0.27 \times 10^{-2}$    & $0.26 \times 10^{-2}$ \\
    $\epsilon_\text{$Z$-peak}$                  & $0.26$          & $0.0062$          & $0.0050$         & $0.035$          & $0.11 \times 10^{-2}$    & $0.14 \times 10^{-3}$ \\
    $\sigma_\text{fiducial} [\text{ab}]$        & $0.58$          & $0.081$           & $0.10$           & $19$             & $0.013$                  & $0.17$ \\
    \hline
  \end{tabular}\\\vspace*{0.5cm}
  \begin{tabular}{c|cccc}
    process                                    & $l^+l^-\nu\nu jj$     & $l^\pm\nu jjj$          & $l^+l^-W^\pm j$          & $l^+l^- jj$              \\
    \hline
    \hline
    $\sigma_\text{MC} [\text{ab}]$              & $15 \times 10^6$      & $2.6 \times 10^9$      &  $0.59\times 10^6$      & $69   \times 10^6$    \\
    $\epsilon_\text{four leptons}$              & $0.81 \times 10^{-5}$ & $0.45 \times 10^{-8}$   &  $0.13 \times 10^{-2}$  & $0.18 \times 10^{-5}$    \\
    $\epsilon_{++--}$                           & $0.45 \times 10^{-5}$ & $0.18 \times 10^{-8}$   &  $0.68 \times 10^{-3}$  & $0.93 \times 10^{-6}$    \\
    $\epsilon_\text{fiducial}$                  & $0.37 \times 10^{-5}$ & $0.87 \times 10^{-9}$   &  $0.62 \times 10^{-3}$  & $0.69 \times 10^{-6}$    \\
    $\epsilon_\text{$Z$-peak}$                  & $0.19 \times 10^{-6}$ & $0.17 \times 10^{-10}$  &  $0.21 \times 10^{-3}$  & $0.40 \times 10^{-7}$    \\
    $\sigma_\text{fiducial} [\text{ab}]$        & $2.7$                 & $0.039$                &  $130$                  & $2.6$          \\
    \hline
  \end{tabular}\\\vspace*{0.5cm}
  \begin{tabular}{c|ccccccccc}
    $m_{H} [\text{GeV}]$                        & 200      & 300      & 400     & 500     & 600      & 700       & 800     & 900      & 1000   \\
    \hline
    \hline
    $\sigma_\text{MC} [\text{ab}]$              & $205$    & $87$     & $42$    & $22$     & $12.4$   & $7.3$    & $4.6$    & $2.9$   & $1.8$  \\
    $\epsilon_\text{four leptons}$              & $0.71$   & $0.77$   & $0.80$  & $0.83$   & $0.84$   & $0.85$   & $0.86$   & $0.86$  & $0.87$ \\
    $\epsilon_{++--}$                           & $0.71$   & $0.76$   & $0.80$  & $0.82$   & $0.83$   & $0.84$   & $0.85$   & $0.85$  & $0.85$ \\
    $\epsilon_\text{fiducial}$                  & $0.63$   & $0.68$   & $0.71$  & $0.74$   & $0.76$   & $0.77$   & $0.77$   & $0.78$  & $0.79$ \\
    $\epsilon_\text{$Z$-peak}$                  & $0.32$   & $0.40$   & $0.48$  & $0.54$   & $0.58$   & $0.61$   & $0.64$   & $0.66$  & $0.68$ \\
    $\sigma_\text{fiducial} [\text{ab}]$        & $66$     & $36$     & $20$    & $12$     & $7.2$    & $4.5$    & $2.9$    & $1.9$   & $1.2$  \\
    \hline
  \end{tabular}
  \caption{List of background and signal processes considered in this
    study. First row: $\sigma_\text{MC}$ is the cross section at
    generator level. Additional rows: cut efficiencies (subsequent
    efficiencies are included multiplicatively, for more details see
    text). Last row: $\sigma_\text{fiducial}$ is the cross section
    after fiducial cuts as used for the analysis section.}
  \label{tab:xsec_mc}
\end{table}
%


\section{Analysis}
\label{sec:analysis}

To analyse and identify kinematic regimes from which our doubly
charged heavy Higgs (we will just use Higgs as abbreviation from now
on) signal might be extracted, we first need to define a fiducial
region using a set of acceptance cuts. We ask for at least four
leptons in the final state; two of them have positive charge
while the other have negative charge. For the four leptons we require
cuts staggered in $p_T$ as follows
\begin{align}
  p_{T,l}^\text{leading} \ge 20~\text{GeV} \qquad
  p_{T,l}^\text{second}  \ge 18~\text{GeV} \qquad
  p_{T,l}^\text{third}   \ge 15~\text{GeV} \qquad
  p_{T,l}^\text{fourth}  \ge 10~\text{GeV}
  \label{eq:fiducial}
\end{align}
as well as require $|\eta_l|\le2.5$. As many of the background
processes contain resonant $Z$ production we require opposite sign,
same flavor leptons to have an invariant mass above $96$ GeV.
\begin{align}
  m_{e^+e^-|\mu^+\mu^-} \ge 96.0\,\text{GeV}\,.
  \label{eq:zpeak}
\end{align}
which also efficiently removes the photon contribution in these events.
Furthermore, as QCD contributes to our missing energy we use the
anti-$k_T$ algorithm, as implemented in \fj, with radius parameter
$R=0.4$, a minimum transverse momentum $p_T^\text{min}=20$~GeV, and
$|\eta_j|\le4.5$ to obtain well defined jets. The missing energy
vector is then computed from
\begin{align}
  \vec{\slashed{p}}_T &= \sum \limits_{i\,\in\, l\,,j} \vec{p}_{T,i}
  \,
  \label{eq:etmiss}
\end{align}
and, in addition, exposed to Eq.~\eqref{eq:smearing}. In
Tab.~\ref{tab:xsec_mc} we collect the cut efficiencies $\epsilon$ and
cross sections $\sigma$ after applying Eq.~\eqref{eq:fiducial},
Eq.~\eqref{eq:zpeak} and requiring at least one muon.

\subsection{Naive Cut and Count}

When analysing the significance of a possible signal or excluding its
existence with a given level of certainty different test statistics
may be used. In our case we implement a simple counting approach based
on a Poisson distribution. Given a sample of expected signal (S) and
background (B) events, the significance $Z$ is given by
\begin{align}
  Z &= \frac{\text{S}}{\sqrt{\text{S}+\text{B}}}\,.
  \label{eq:sign_def}
\end{align}
Sometimes $\text{S}/\sqrt{\text{B}}$ is used as well. However, we
emphasize that for low background cross sections, as we face in our
study, this leads to wrong conclusions. Take, for example, S~$=1$ and
B~$=0.1$. This would yield a significance of $3$, which is clearly an
overestimate when only on event in total is expected. The consequence
of Eq.~\eqref{eq:sign_def} is that one needs to expect at least four
signal events in the analysis region to set an exclusion limit of
$95\%$. As we see, this cannot be possible for the very high mass regime
of our model, even when considering the $3~\text{ab}^{-1}$ high luminosity
LHC option. However, for completeness we will show results up to
$m_{H^{\pm\pm}}=1000$~GeV. 

It is also possible to invert
Eq.~\eqref{eq:sign_def}.
\begin{align}
  \mathcal{L} &= \frac{ Z^2 \left( \sigma_\text{S} + \sigma_\text{B}
    \right)}{ \sigma_\text{S}^2}\,.
  \label{eq:target_lumi}
\end{align}
For given signal and background cross sections setting $Z\equiv 2$
yields the expected luminosity needed for an exclusion at $95\%$
confidence level.

\begin{figure}[t]
  \centering
  \includegraphics[width=0.23\textwidth]{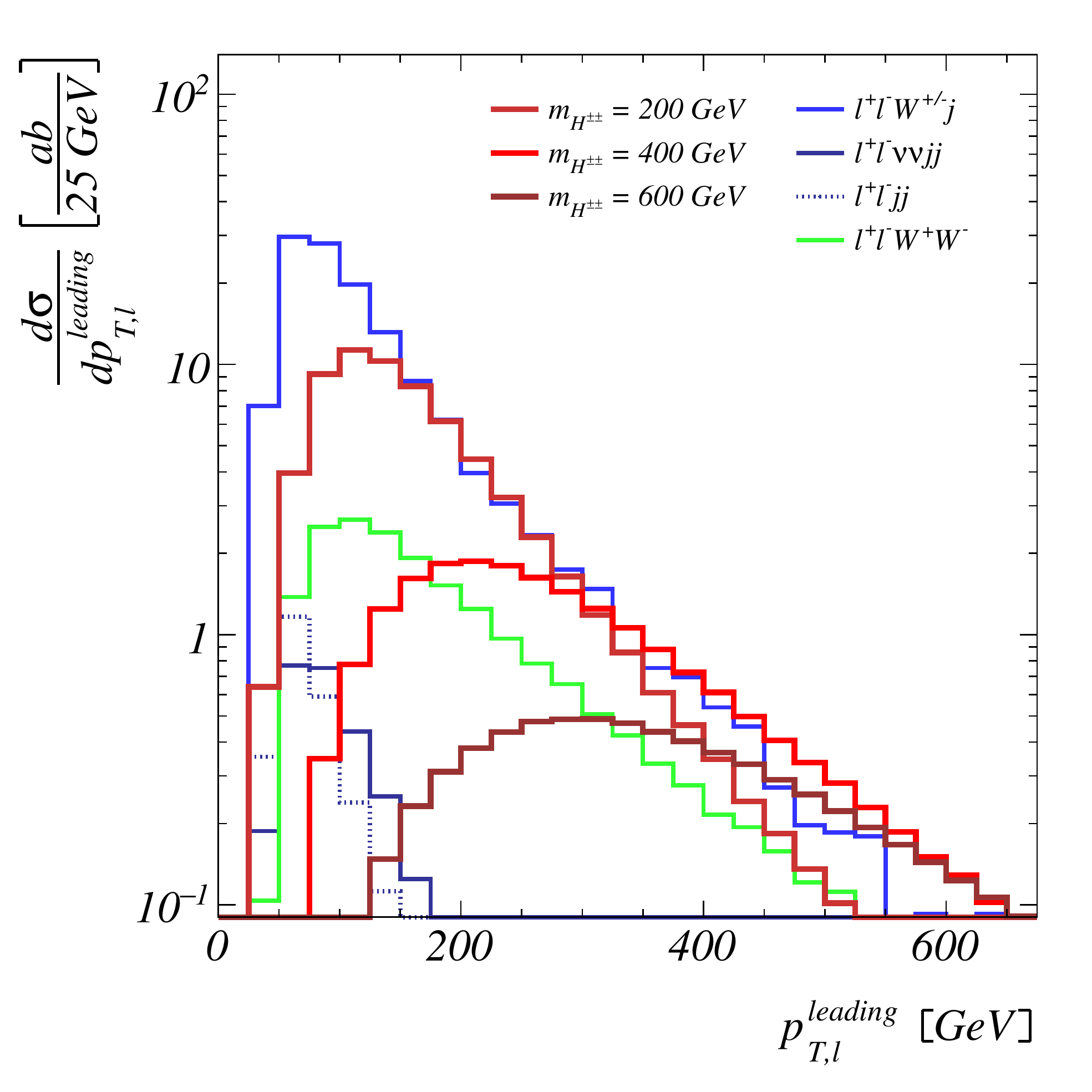}\hspace*{2mm}
  \includegraphics[width=0.23\textwidth]{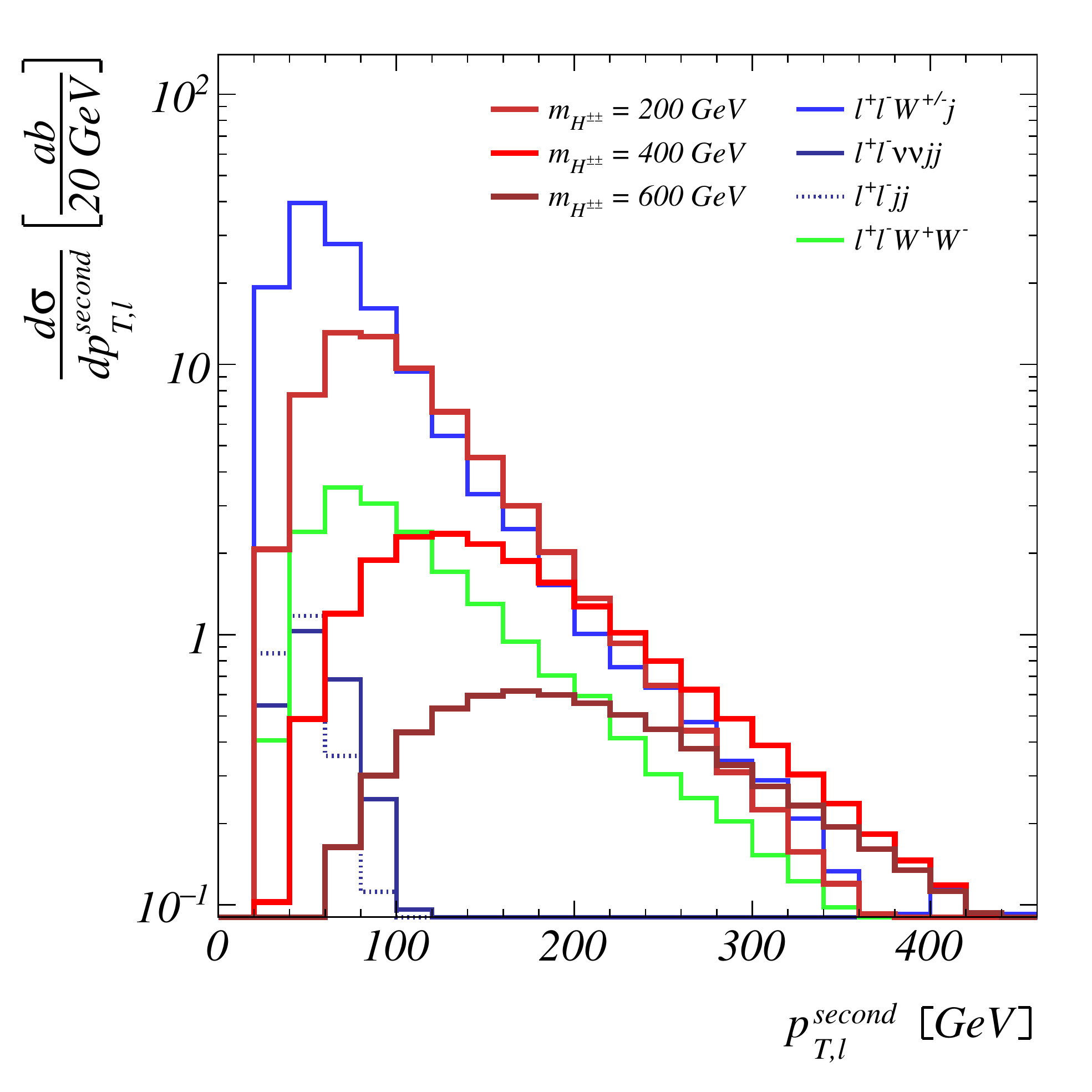}\hspace*{2mm}
  \includegraphics[width=0.23\textwidth]{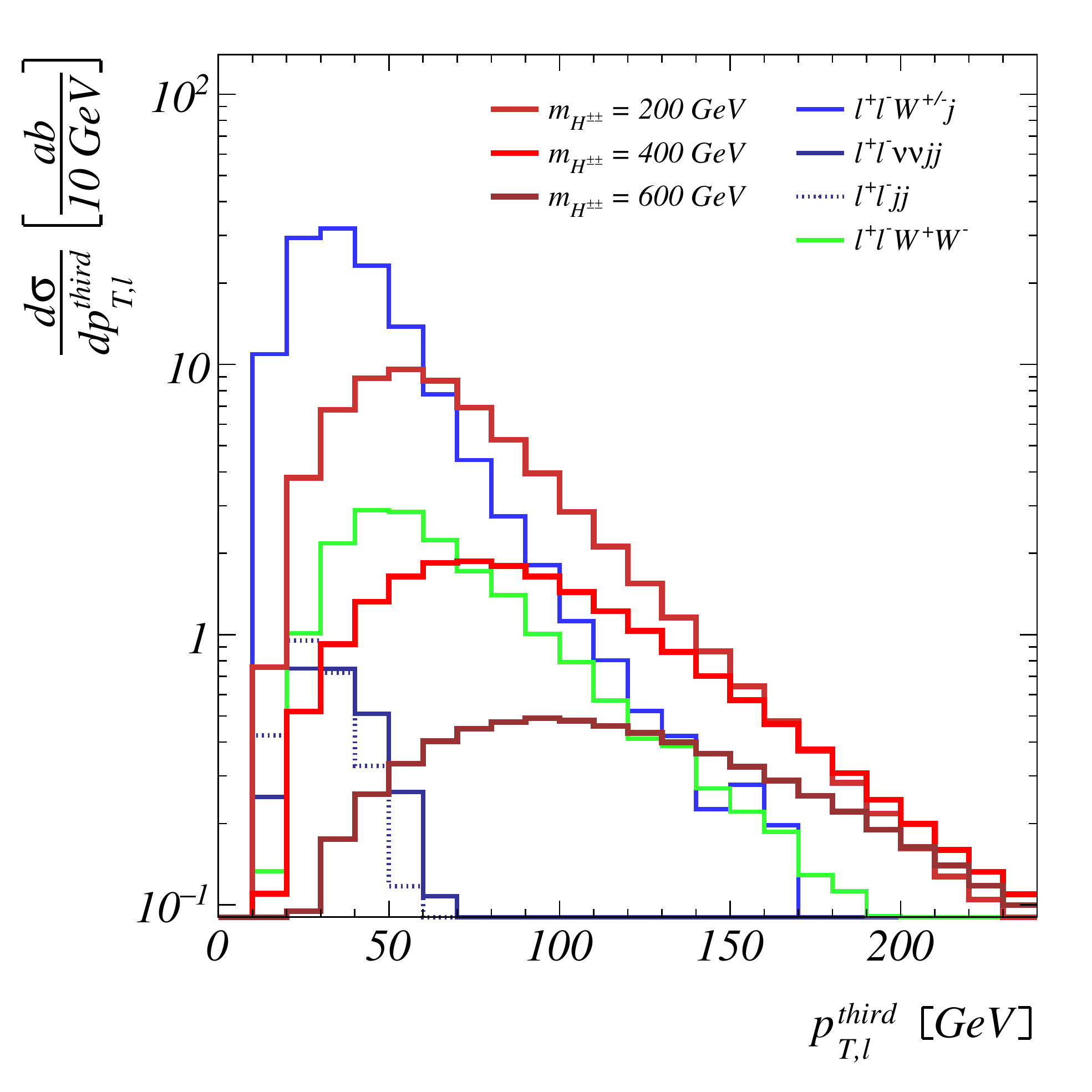}\hspace*{2mm}
  \includegraphics[width=0.23\textwidth]{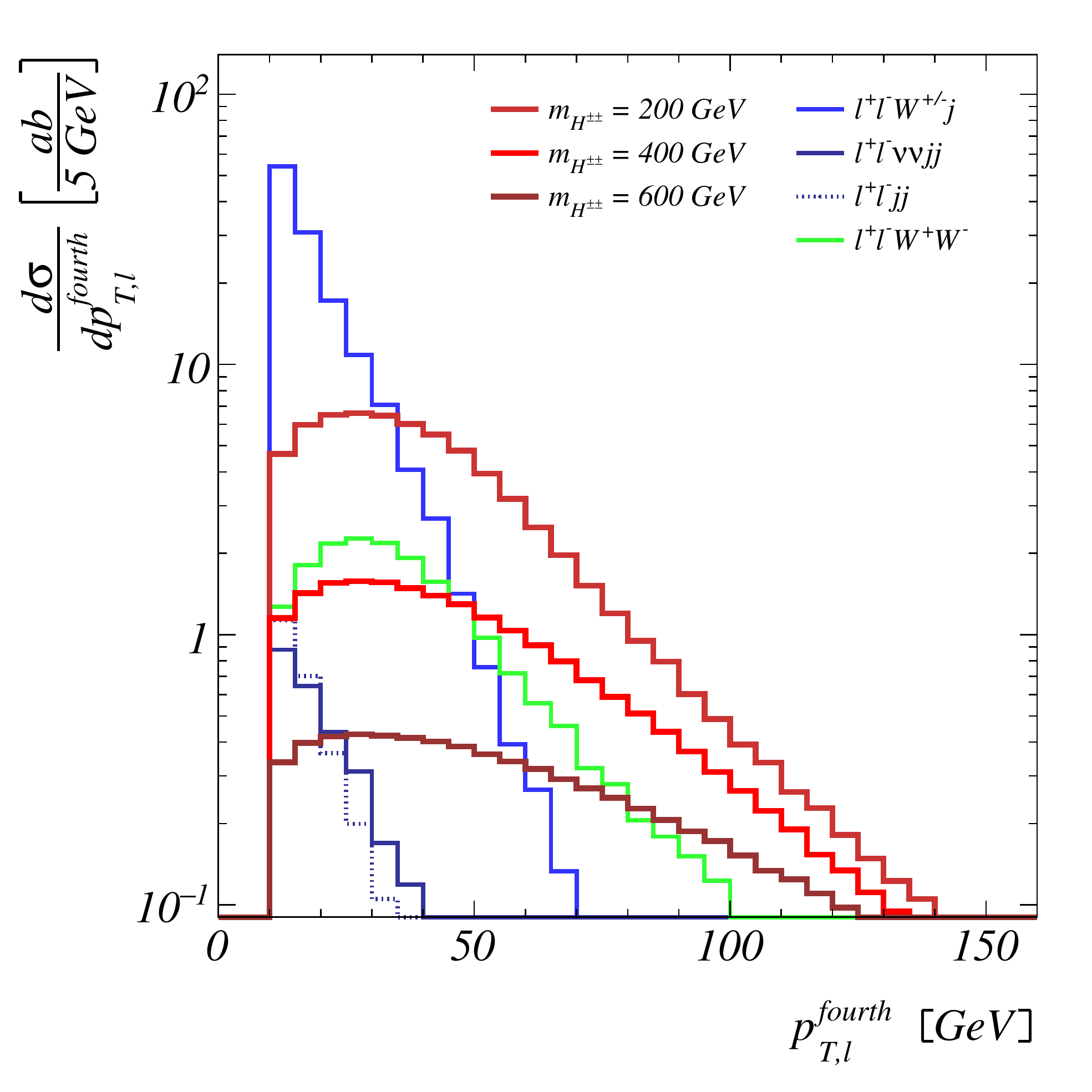}
  \caption{Transverse momentum of the leptons sorted by $p_T$ after
    Eq.~\eqref{eq:fiducial}. Red: signal with $m_{H^{\pm\pm}} \in
    [200,400,600]$~GeV. Blue: Backgrounds from jets faking
    electrons. Green: irreducible backgrounds.}
  \label{fig:pts}
\end{figure}

For this first part of our analysis we assume the branching ratio
Higgs to $WW$ to be unity. Under this assumption we check observables
which should yield discriminative power one by one. As the dominant
backgrounds are driven by QCD radiation mistaken for electrons or an
electro-weak topology with masses well below 100 GeV while our signal
stems from a heavy decay, the $p_T$ spectra should provide a sensitive
discriminant. In Fig.~\ref{fig:pts} we show the $p_T$ for all
backgrounds and a sub-sample of the Higgs mass parameter space. Note
that backgrounds not listed in the legend of the plot are too small to
be considered.
We find that the QCD-induced backgrounds fall more steeply than the
signal, see e.g. $p_{T,l}^\text{third}$ of $l^+l^-jj$ or
$p_{T,l}^\text{fourth}$ of $l^+l^-W^\pm j$.

We note that despite the fact that the $1/p_T$ divergence of QCD shows
itself in $p_{T,l}^\text{third}$ and $p_{T,l}^\text{fourth}$, these do not
yield much discriminating power. Therefore, for the further analysis steps, we
fix $p_{T,l}^\text{third}=20$~GeV, $p_{T,l}^\text{fourth}=15$~GeV and
manipulate the leading leptons' transverse momenta to obtain a signal efficiency of
$\epsilon_\text{S}=0.9$. We tabulate the $p_T$ cuts needed for this
together with the resulting background cross section $\sigma_\text{B}$
in Tab.~\ref{tab:pt_point}.
\begin{table}[!b]
  \renewcommand{\arraystretch}{1.5}
  \centering
  \begin{tabular}{ll|ccccccccc}
    $m_{H}~[\text{GeV}]$           &~                & 200   & 300    & 400    & 500    & 600   & 700    & 800   & 900    & 1000   \\
    \hline
    \hline
     cut on $p_{T,l}^\text{leading}$ & $[\text{GeV}]$   & 70   & 95     & 120    & 140    & 160   & 180    & 200   & 220    & 240    \\
     cut on $p_{T,l}^\text{second} $ & $[\text{GeV}]$   & 35   & 40     & 40     & 40     & 40    & 40     & 60    & 60     & 70    \\
     \hline
     $\sigma_\text{B}\,[\text{ab}]$ &~                & 74   & 54      & 40    & 31     & 24    & 20     & 15    & 13     & 9.9  \\
    \hline
  \end{tabular}
  \caption{Transverse momentum cut on the two leading leptons to obtain a
    $\epsilon_\text{S}=0.9$ working point and corresponding background
    cross section $\sigma_\text{B}$ for different mass parameters
    $m_{H^{\pm\pm}}$.}
  \label{tab:pt_point}
\end{table}

The leptonic decay of the $W$'s in the signal channel results in
missing energy. Furthermore, we expect the jet radiation pattern to be
sensitive to the presence of the electroweak
decay~\cite{Englert:2011cg,Gerwick:2011tm}. However, we find that the
jet spectrum after $p_T$ cuts is not sensitive to the decay
pattern. At least for the reducible backgrounds, the jet number
spectrum does not resemble the correct starting point to count the number of
jets following the strategy of~\cite{Gerwick:2012hq}. It is known that the jet
spectrum poses a good discriminator before restrictive $p_T$ cuts
sculpt the distribution's shape~\cite{Englert:2011pq}, hence, it is plausible
that it loses its sensitivity in the present case through decays. Also, only
one out of two leading backgrounds stems from QCD; the $l^+l^-W^+W^-$
channel is expected to have a similar radiation pattern as our
signal. We therefore do not use the number of jets to construct analysis
cuts; this means we also do not impose a jet veto. However, if a jet
veto should become necessary due to experimental reasons beyond the scope
of this study, our analysis shows that this should not decrease the signal efficiency too much. 
\begin{figure}[t]
  \centering
  \includegraphics[width=0.31\textwidth]{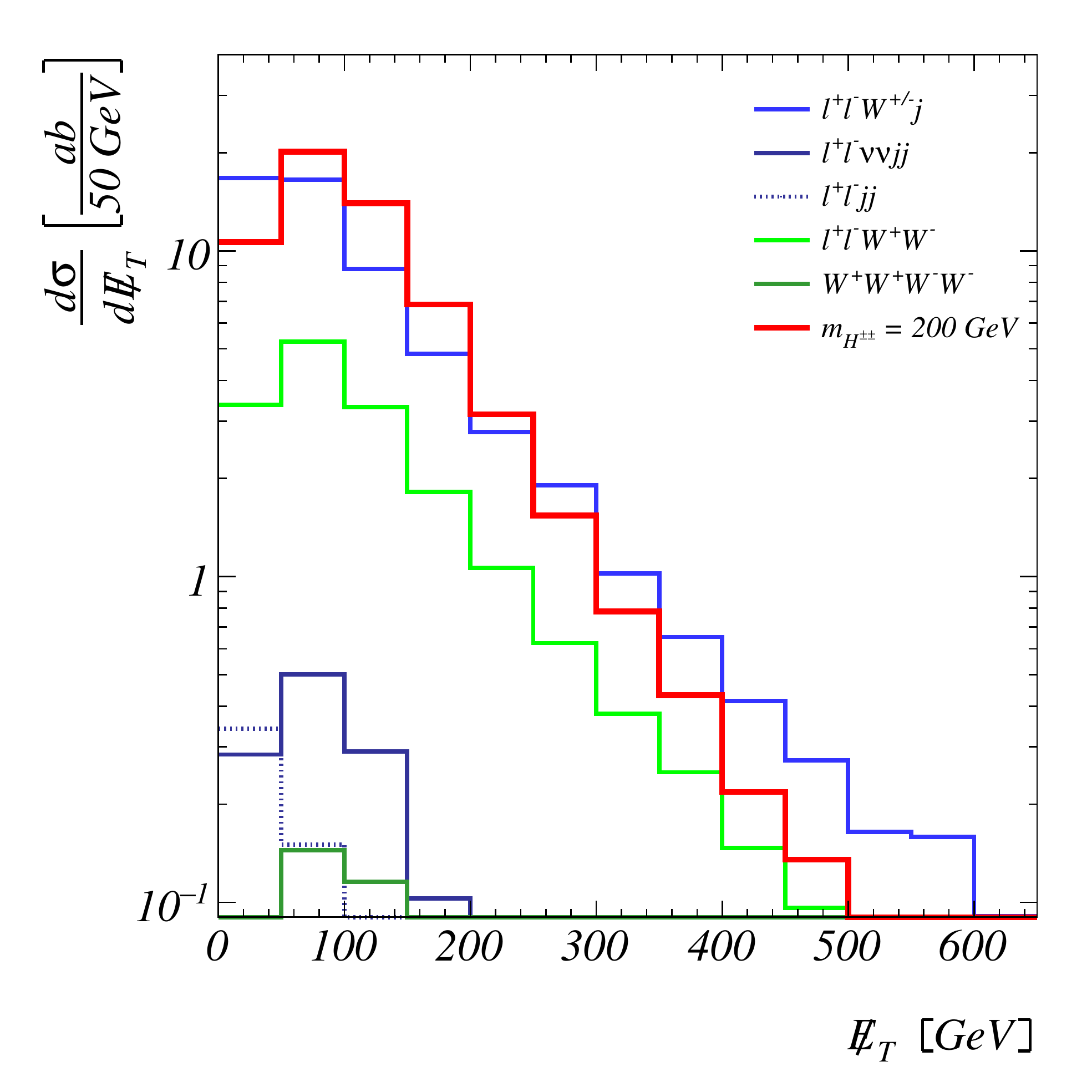}\hspace*{4mm}
  \includegraphics[width=0.31\textwidth]{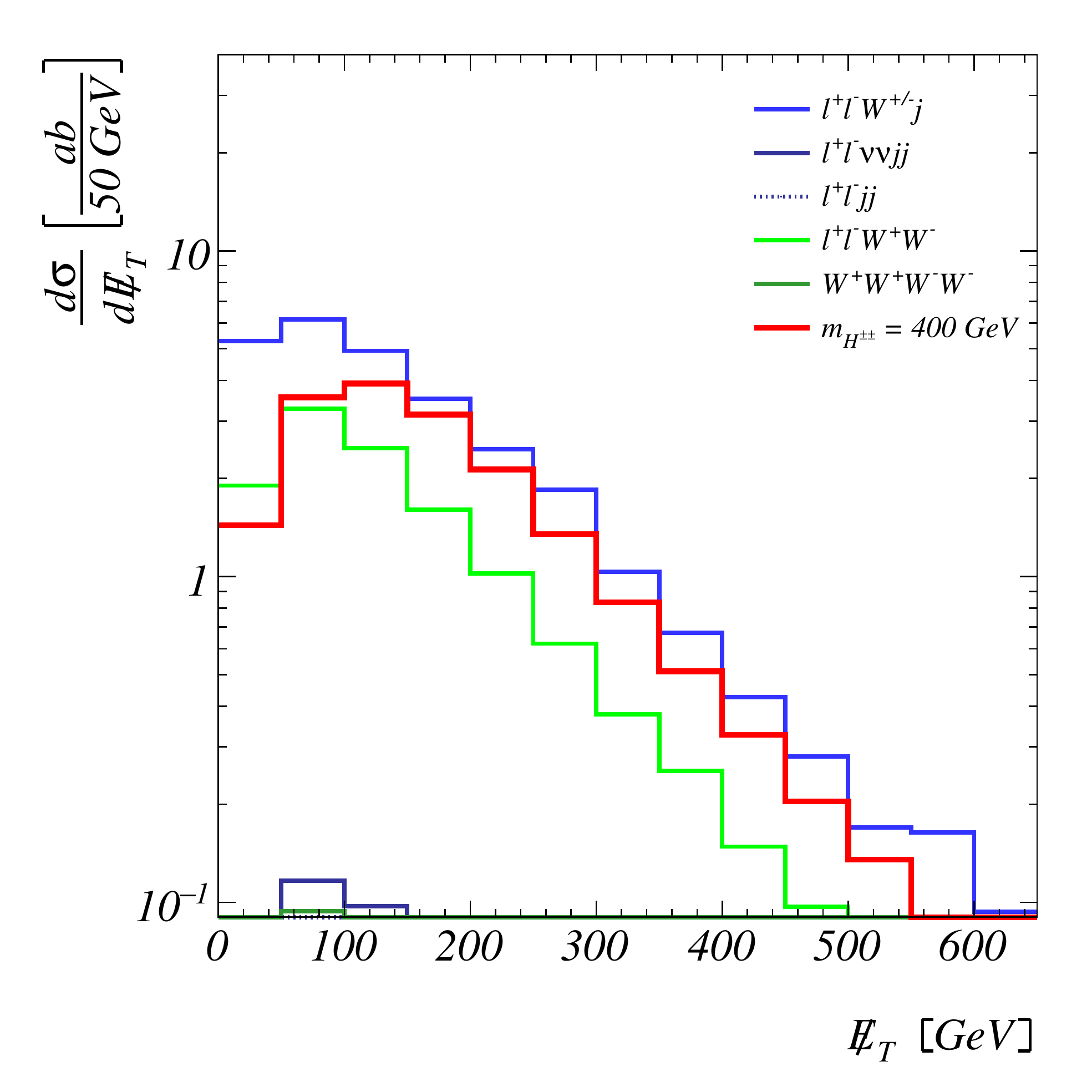}\hspace*{4mm}
  \includegraphics[width=0.31\textwidth]{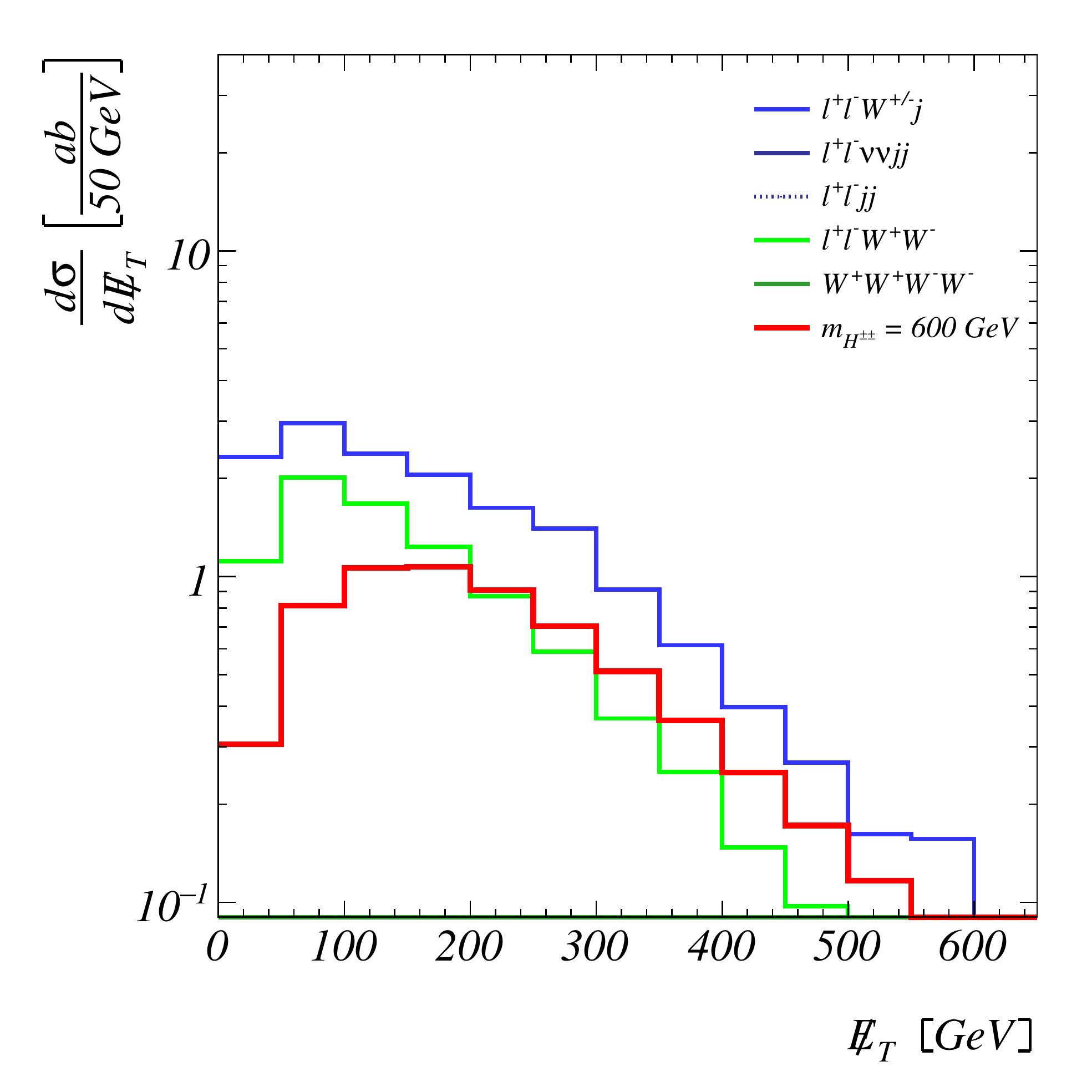}
  \caption{Missing energy spectrum after cuts outlined in
    Tab.~\ref{tab:pt_point}. Red: from left to right $m_{H^{\pm\pm}}
    \in [200,400,600]$~GeV. Blue: backgrounds from jets faking
    electrons. Green: irreducible backgrounds.}
  \label{fig:etmiss}
\end{figure}

The missing energy is not
very sensitive to the signal process after applying the cuts outlined
in Tab.~\ref{tab:pt_point}, as can be seen from Fig.~\ref{fig:etmiss}.
The $p_T$ cuts force the QCD radiation
into a regime where also jets produce a fair amount of missing
energy. Thus we propose only a mild cut of
\begin{align}
  \slashed{E}_T > 20.0 ~\text{GeV}\,.
  \label{eq:etmiss_cut}
\end{align}

\begin{figure}[t]
  \centering
  \includegraphics[width=0.31\textwidth]{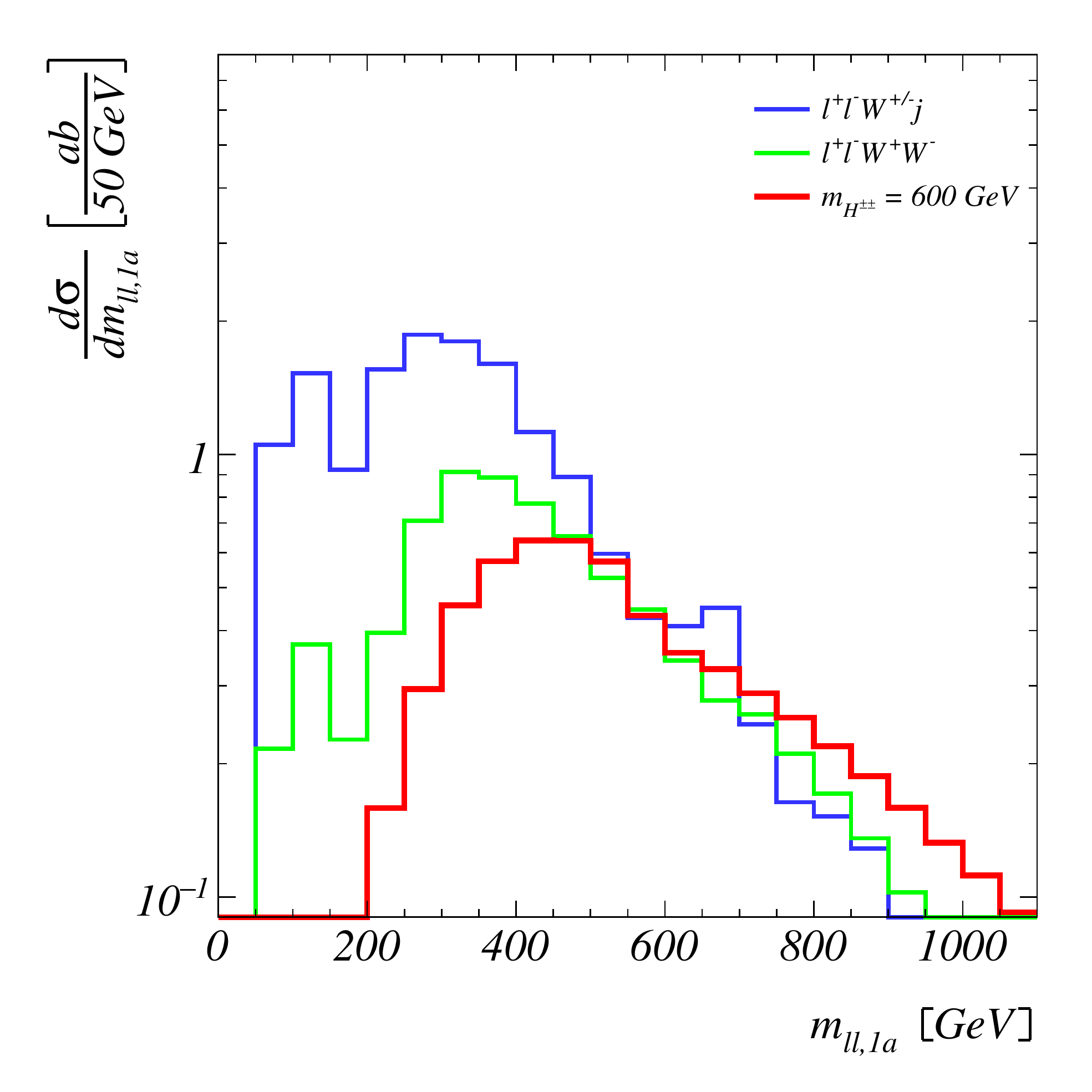}\hspace*{4mm}
  \includegraphics[width=0.31\textwidth]{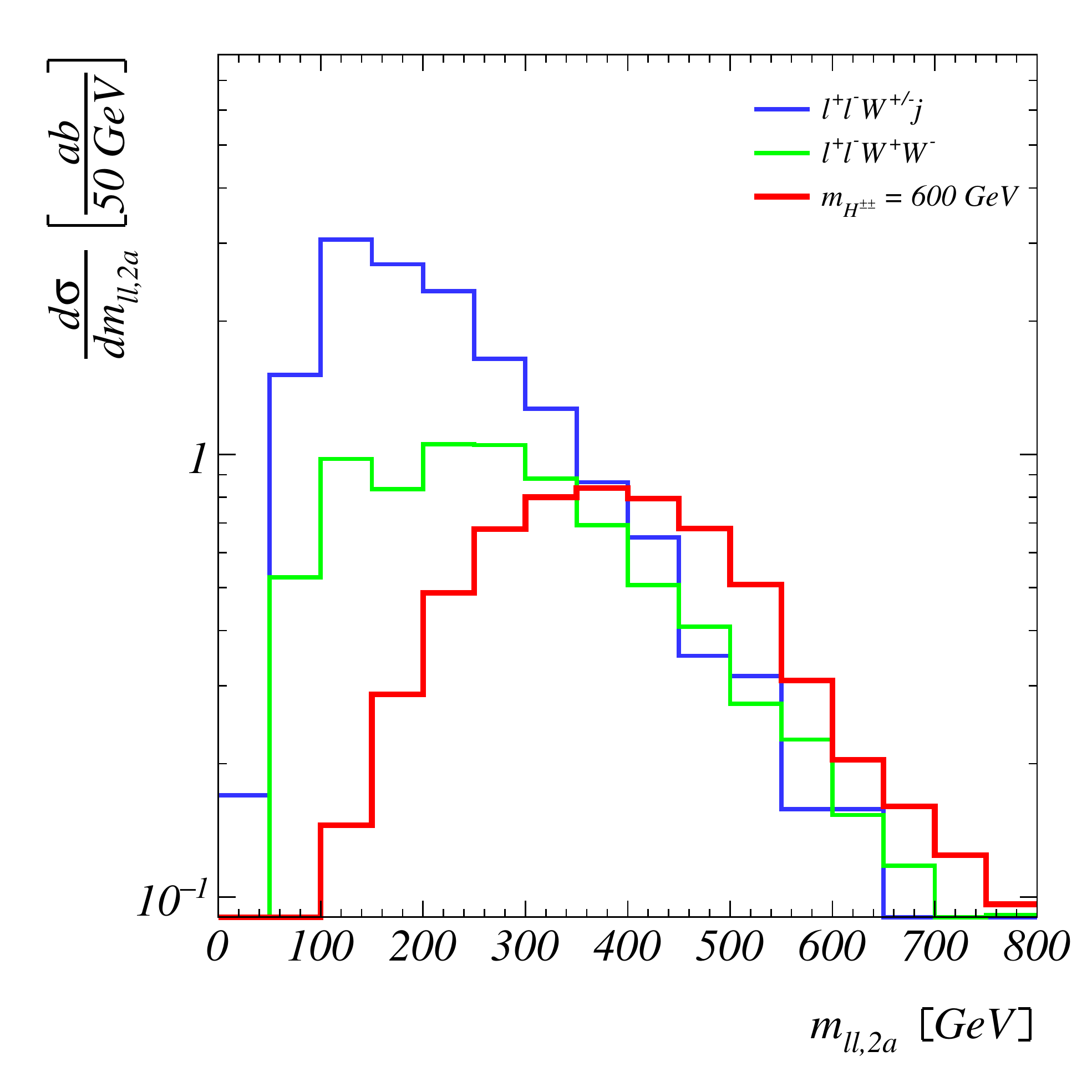}\hspace*{4mm}
  \includegraphics[width=0.31\textwidth]{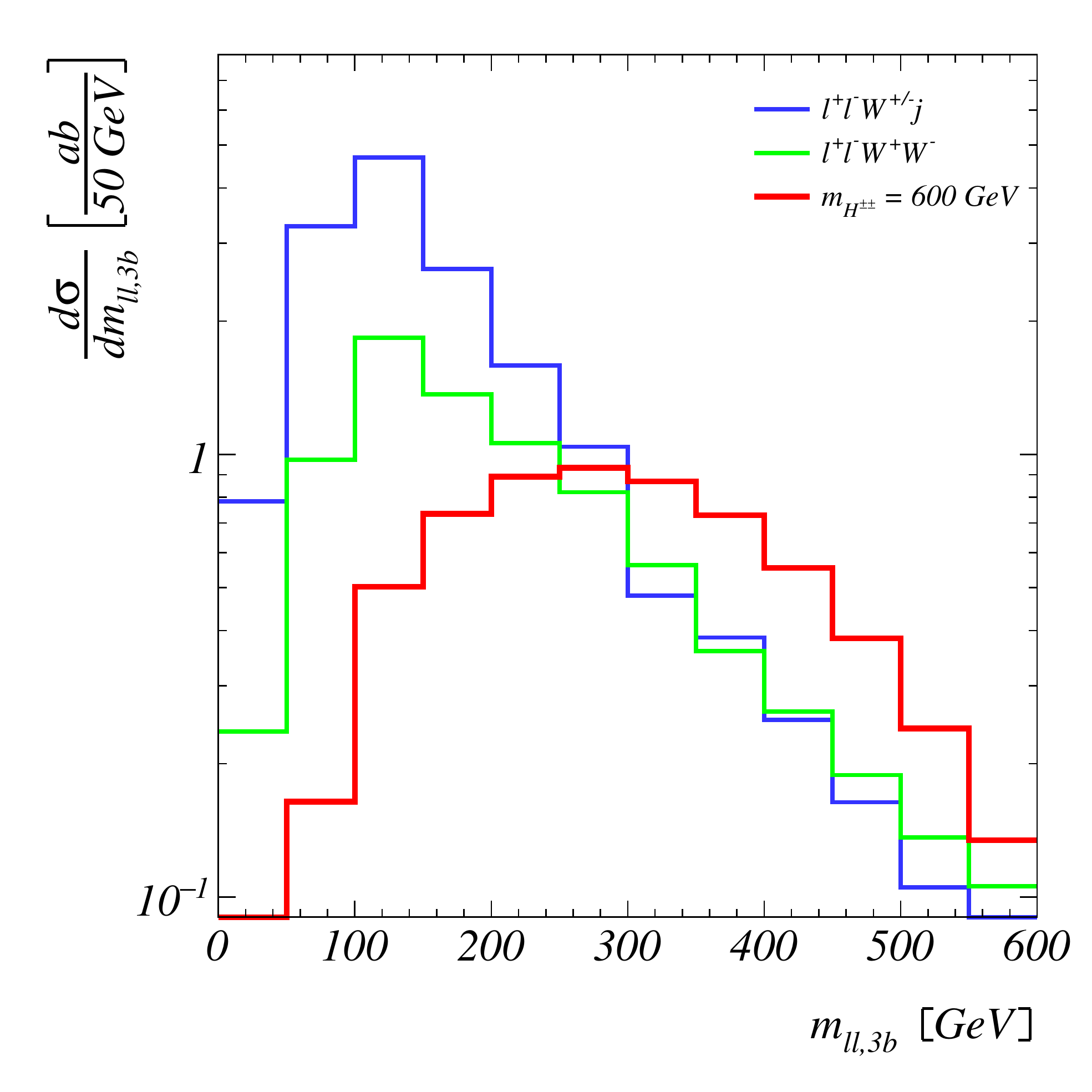}
  \caption{Lepton pair invariant mass reconstruction after cuts
    outlined in Tab.~\ref{tab:pt_point} and
    Eq.~\eqref{eq:etmiss}. Red: signal with
    $m_{H^{\pm\pm}}=600$~GeV. Blue: backgrounds from jets faking
    electrons. Green: irreducible backgrounds. From left to right:
    $m_{ll,1a}$, $m_{ll,2a}$, and $m_{ll,3b}$ (for details see text
    and Eq.~\eqref{eq:mz_cuts})}
  \label{fig:mz}
\end{figure}

There is another feature we can use to enhance our signal to
background ratio: Many of the backgrounds contain either resonant $Z$
boson contributions or have no resonance connecting two of the
leptons. Thus, we construct all possible invariant masses build out of
two lepton four momenta added to each other. This yields a total of
six observables.
\begin{alignat}{4}
  m_{ll,1a} &= \text{mass}\left( \, p_l^\text{leading} + p_l^\text{second} \,\right) &\qquad\text{and}&\qquad
  m_{ll,1b} &= \text{mass}\left( \, p_l^\text{third  } + p_l^\text{fourth} \,\right) \,,\notag\\
  m_{ll,2a} &= \text{mass}\left( \, p_l^\text{leading} + p_l^\text{third } \,\right) &\qquad\text{and}&\qquad
  m_{ll,2b} &= \text{mass}\left( \, p_l^\text{second } + p_l^\text{fourth} \,\right) \,,\notag\\
  m_{ll,3a} &= \text{mass}\left( \, p_l^\text{leading} + p_l^\text{fourth} \,\right) &\qquad\text{and}&\qquad
  m_{ll,3b} &= \text{mass}\left( \, p_l^\text{second } + p_l^\text{third } \,\right) \,.
  \label{eq:mz_cuts}
\end{alignat}
Interestingly, only the combinations $1a,2a$ and $3b$ yield any
additional discriminating features. We show examples for some of these
for $m_{H^{\pm\pm}}=600$~GeV in Fig.~\ref{fig:mz}.

When designing the cut flow for this particular signal search, we have to balance two
effects. On the one hand, we have to ensure a strong background rejection
while, on the other hand, we need to ensure that the already low signal yield does not become further suppressed. 
For the low mass points, we find that signal and
background have very similar kinematic features. We therefore suggest to not cut
on the invariant lepton masses is applied when studying the parameter space below
$m_{H^{\pm\pm}}=300$~GeV. For the rest of the parameter space we again
aim for $\epsilon_\text{S}=0.9$. We show our cut flow in
Tab.~\ref{tab:mz_cuts}.
\begin{table}[b]
  \renewcommand{\arraystretch}{1.5}
  \centering
  \begin{tabular}{l|ccccccccc}
    $m_{H}~[\text{GeV}]$                      & 200   & 300    & 400    & 500    & 600   & 700   & 800 & 900  & 1000   \\
    \hline
    \hline
    $\slashed{E}_T~[\text{GeV}]$             & 20   & 20      & 20    & 20     & 20    & 20    & 20   & 20   & 20 \\
    \hline
    $m_{Z,1a}~[\text{GeV}]$                   &   ~   & 65      & 130   & 125    & 125   & 170   & 215  & 165  & 225 \\
    $m_{Z,2a}~[\text{GeV}]$                   &   ~   & 110     & 135   & 125    & 160   & 145   & 210  & 185  & 240 \\
    $m_{Z,3b}~[\text{GeV}]$                   &   ~   & 25      & 65    & 110    & 120   & 140   & 130  & 165  & 155 \\
    \hline
    $R_{l^+l^+}^\text{max}$                    & 2.2   & 2.3     & 2.8   & 2.8     & 2.8  & 2.8   & 2.8   & 2.8  & 2.8 \\
    \hline
     $\sigma_\text{S}\left[\text{ab}\right]$  & 51    & 25      & 15    & 8.8    & 5.2   & 3.2  & 2.0  & 1.3  & 0.8   \\
     $\sigma_\text{B}\left[\text{ab}\right]$  & 24    & 16      & 14    & 8.8    & 6.1   & 4.4  & 3.4  & 2.6  & 2.0   \\
    \hline
  \end{tabular}
  \caption{Cut flow optimised for different Higgs mass scenarios. Last
    rows: Signal cross section $\sigma_\text{S}$ and background cross
    section $\sigma_\text{B}$ for different mass parameters
    $m_{H^{\pm\pm}}$ after the cuts stated here, see
    Eq.~\eqref{eq:etmiss}, Eq.~\eqref{eq:mz_cuts}, and
    Eq.~\eqref{eq:dr}.}
  \label{tab:mz_cuts}
\end{table}

\begin{figure}[t]
  \centering
  \includegraphics[width=0.31\textwidth]{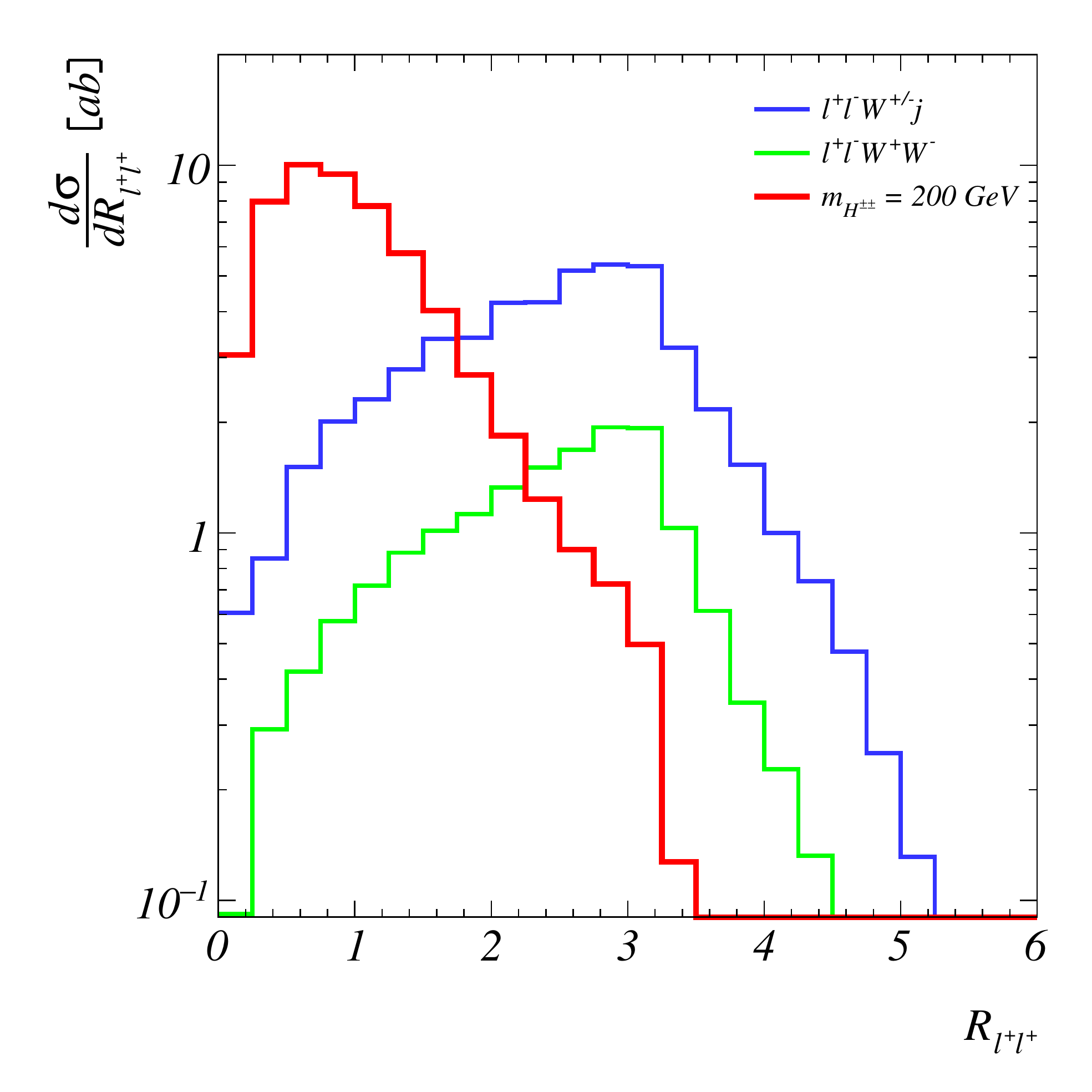}\hspace*{4mm}
  \includegraphics[width=0.31\textwidth]{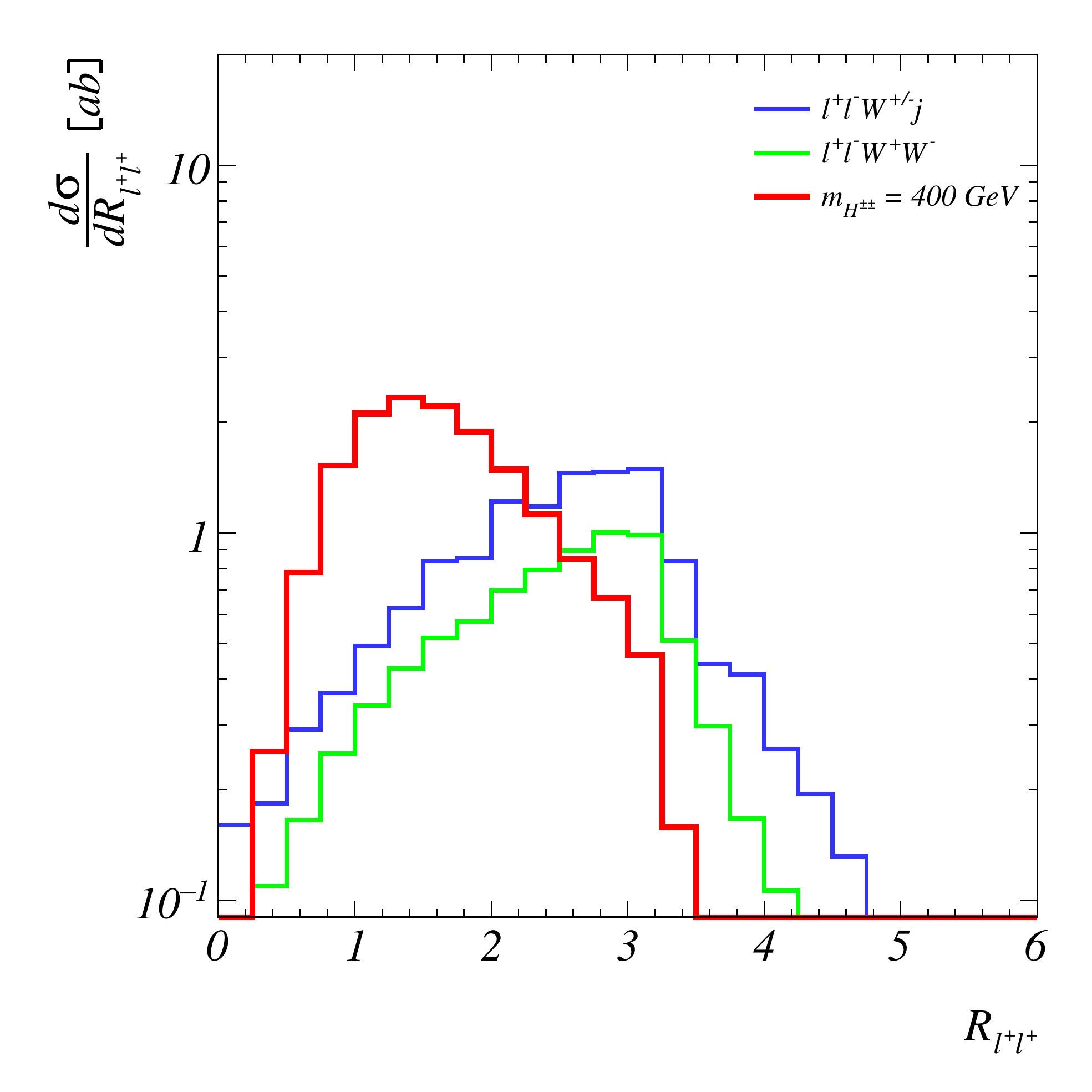}\hspace*{4mm}
  \includegraphics[width=0.31\textwidth]{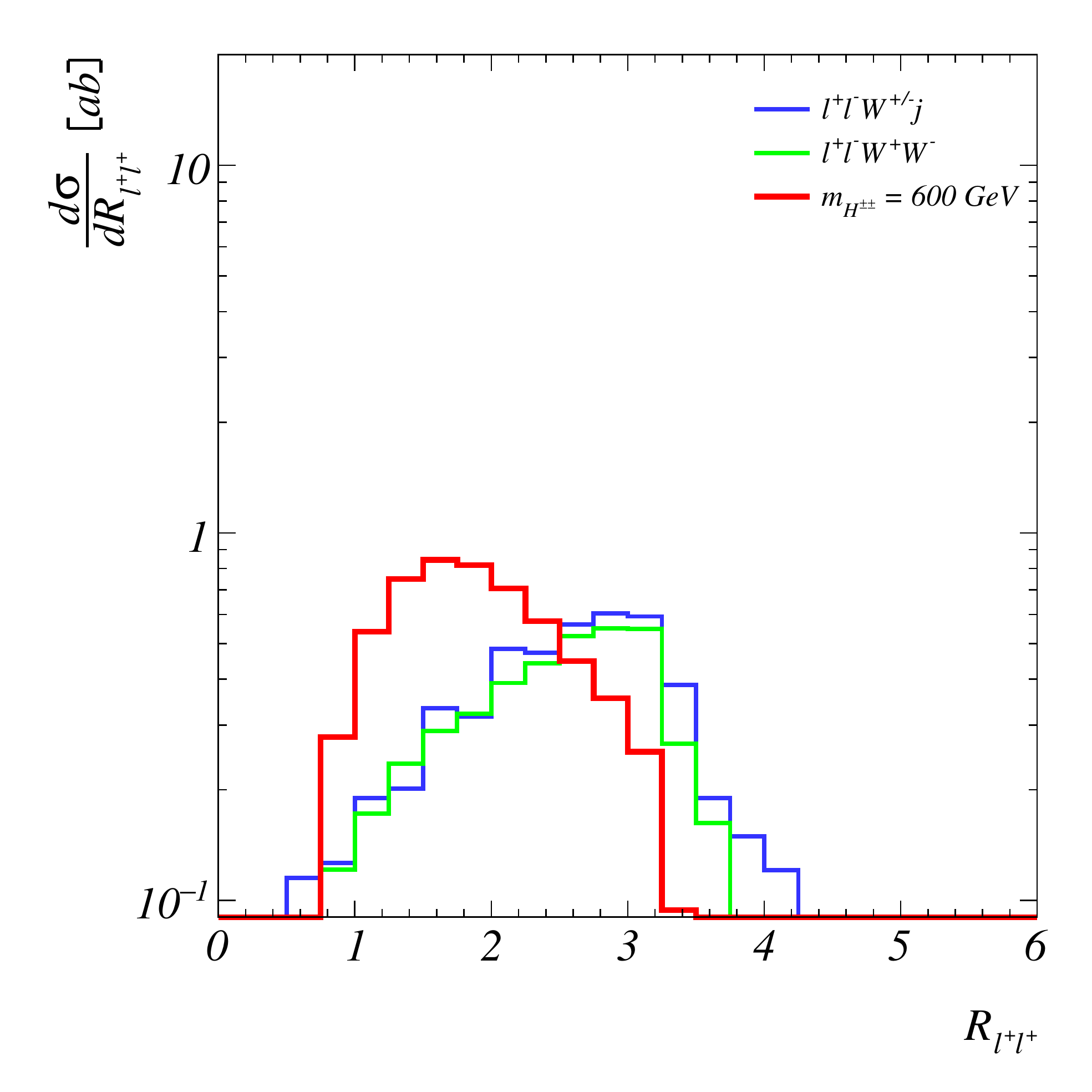}
  \caption{Radial separation of the same sign lepton pair as defined
    in Eq.~\eqref{eq:dr} (positive choice). Red: from left to right
    $m_{H^{\pm\pm}} \in [200,400,600]$~GeV. Blue: backgrounds from
    jets faking electrons. Green: irreducible backgrounds.}
  \label{fig:dr}
\end{figure}

Since the Higgs pairs are produced back-to-back, we expect the decay products of the Higgs bosons will dominantly fall into
the same detector hemispheres
\begin{align}
  R_{l^\pm l^\pm} &= \sqrt{ \Delta\eta^2_{l^\pm l^\pm} +
    \Delta\phi^2_{l^\pm l^\pm} }\,,
  \label{eq:dr}
\end{align}
where $\Delta\eta$ denotes the difference in rapidity between the two
leptons while $\Delta\phi$ is their azimuthal separation. In
Fig~\ref{fig:dr} we show the radial distance between the same charge
leptons. Indeed, these requirements are a powerful discriminator and we suggest cuts on
the maximum value in Tab.~\ref{tab:mz_cuts}.

\begin{figure}[t]
  \centering
  \includegraphics[width=0.31\textwidth]{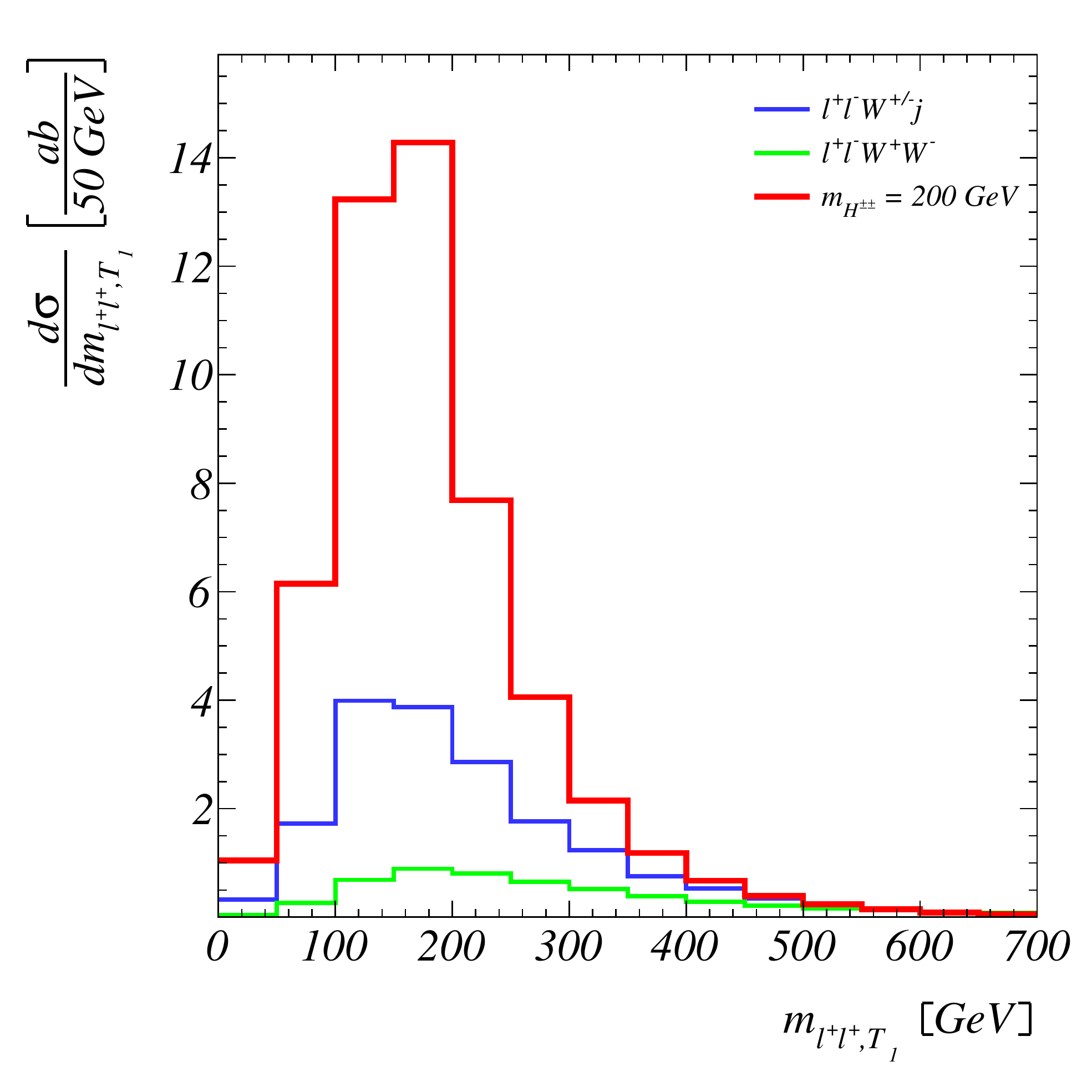}\hspace*{4mm}
  \includegraphics[width=0.31\textwidth]{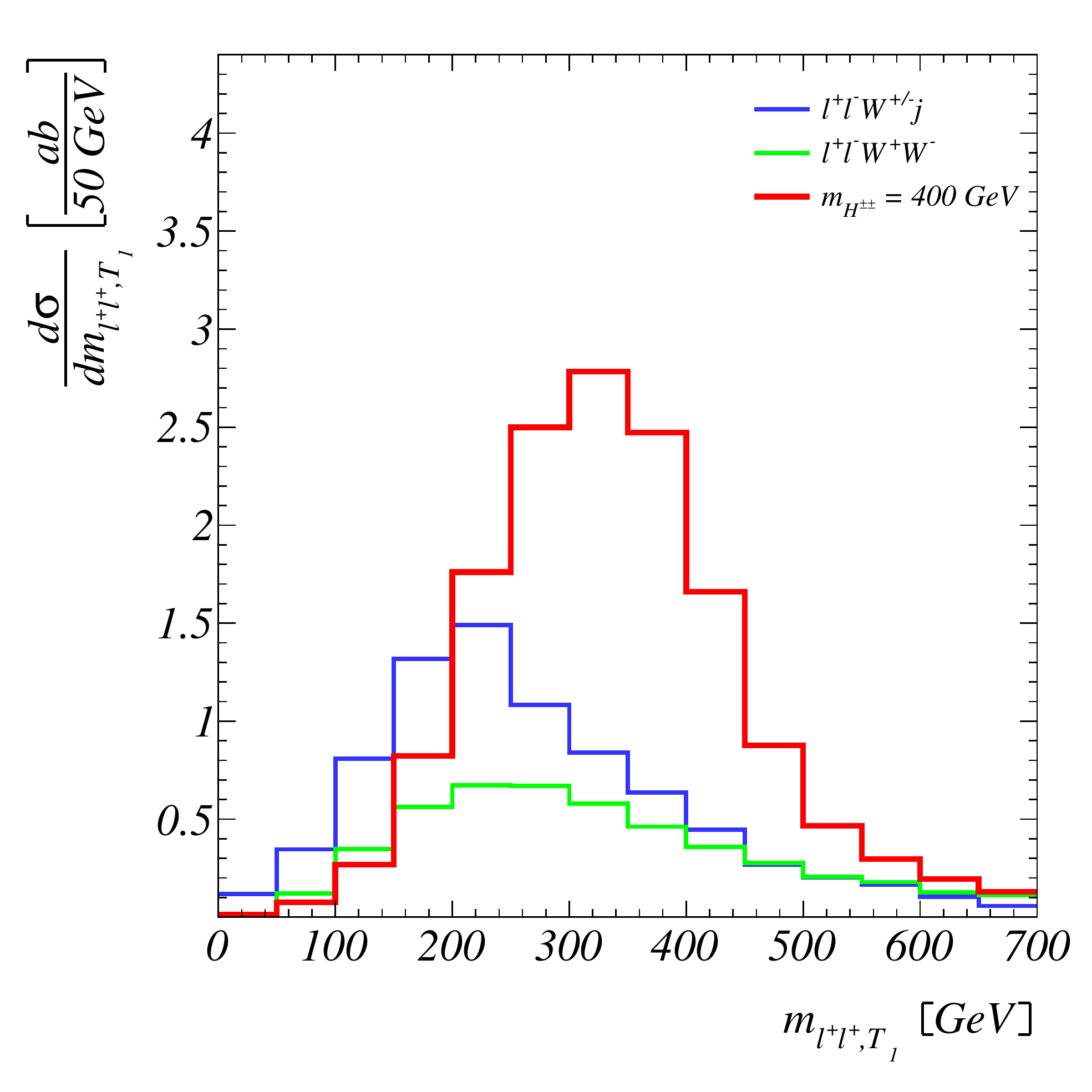}\hspace*{4mm}
  \includegraphics[width=0.31\textwidth]{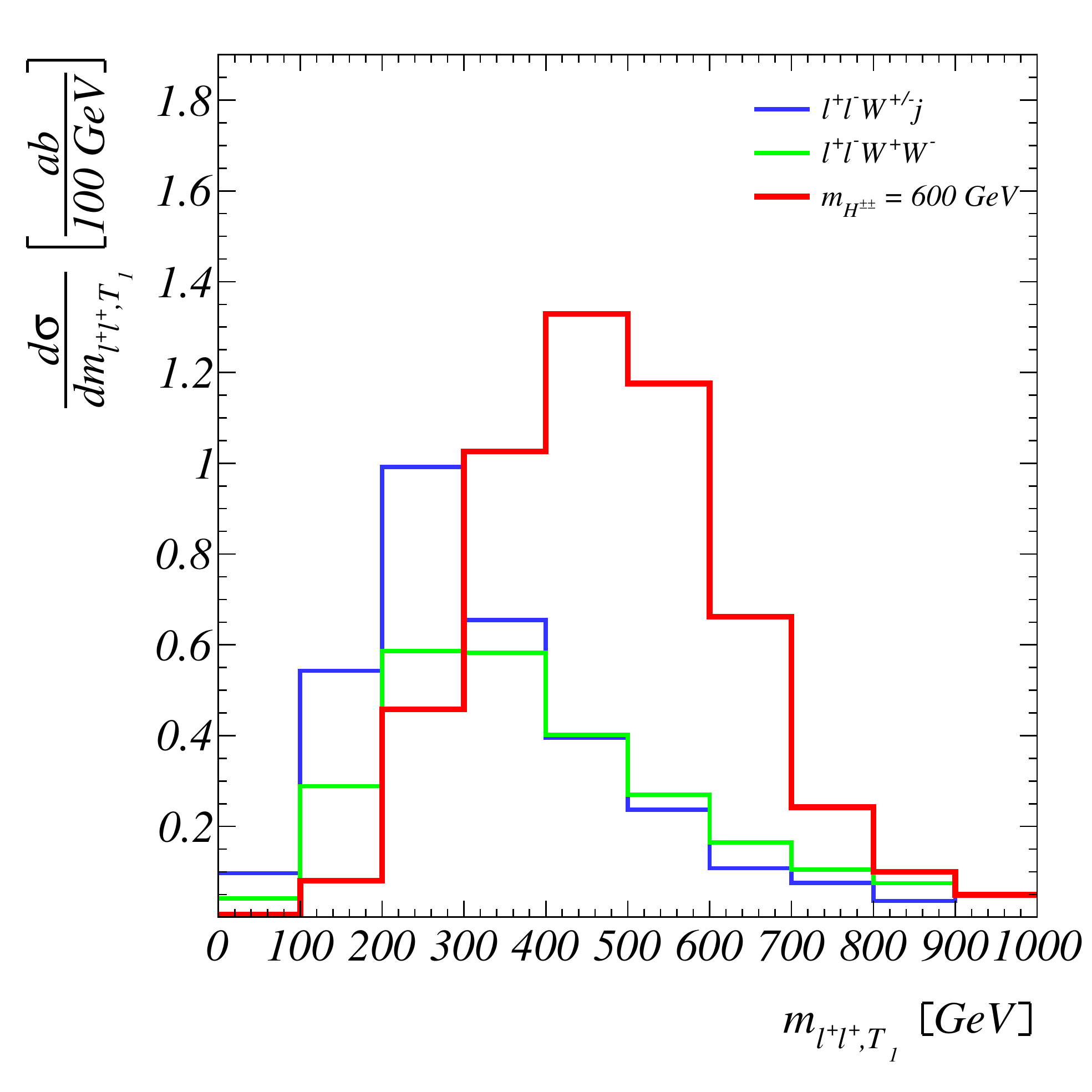}
  \caption{Transverse mass after cuts outlined in
    Tab.~\ref{tab:mz_cuts}. Red: from left to right $m_{H^{\pm\pm}}
    \in [200,400,600]$~GeV. Blue: backgrounds from jets faking
    electrons. Green: irreducible backgrounds.}
  \label{fig:mt1}
\end{figure}

As we produce both Higgs bosons on-shell, the invariant mass should
naively be a very good discriminator. However, since there is a fair
amount of missing energy in this particular final state, we focus
instead on selections of adapted definitions of the transverse mass
$m_T$. There are different definitions of $m_T$ depending on the
phenomenological circumstances.  For instance,
Ref.~\cite{Plehn:1999xi} defines
\begin{align}
  m^2_{l^\pm l^\pm,T_1} &= \Big[ \sqrt{ p_{l^\pm l^\pm,T}^2 +
          m_{l^\pm l^\pm}^2 } + \slashed{p}_T \Big]^2 - \Big[
        p_{l^\pm l^\pm,x} + \slashed{p}_x \Big]^2 - \Big[
        p_{l^\pm l^\pm,y} + \slashed{p}_y \Big]^2 \,.
  \label{eq:mt1}
\end{align}
and Ref.~\cite{Kauer:2000hi}
\begin{align}
  m^2_{l^\pm l^\pm,T_2} &= \Big[ p_{l^\pm l^\pm,T} + \slashed{p}_T
    \Big]^2 - \Big[ p_{l^\pm l^\pm,x} + \slashed{p}_x \Big]^2 -
  \Big[ p_{l^\pm l^\pm,y} + \slashed{p}_y \Big]^2 \,.
  \label{eq:mt2}
\end{align}
The only difference between these definitions is the inclusion of the
invariant lepton mass. For the case of a heavy resonance, as studied
here, the former one yields better
discrimination\footnote{Eq.~\eqref{eq:mt2} pushes both signal and
  background into the low mass region.}. Therefore, we show
Eq.~\eqref{eq:mt1} for a sample of Higgs mass points in
Fig.~\ref{fig:mt1}.

Using this $m_T$ as final discriminant, we scan along the $m_T$ axis to minimise the amount of
luminosity needed for a $95\%$ exclusion as computed from
Eq.~\eqref{eq:target_lumi}. We collect our results in
Tab.~\ref{tab:mt_excl}.
\begin{table}[b]
  \renewcommand{\arraystretch}{1.5}
  \centering
  \begin{tabular}{l|ccccccccc}
    $m_{H}~[\text{GeV}]$                                           & 200   & 300    & 400   & 500  & 600   & 700  & 800  & 900  & 1000   \\
    \hline
    \hline
    cut on $m_T~[\text{GeV}]$                                      & 2.0   & 2.0    & 173   & 234  & 280   & 341  & 367  & 457   & 488   \\
    \hline
    $\sigma_\text{S}\left[\text{ab}\right]$                         & 51    & 25     & 14    & 8.1  & 4.8   & 2.9  & 1.9  & 1.1   & 0.71  \\
    $\sigma_\text{B}\left[\text{ab}\right]$                         & 24    & 16     & 11    & 6.1  & 3.8   & 2.3  & 1.8  & 1.1   & 0.79  \\
    \hline
    $\mathcal{L} \left[\text{ab}^{-1}\right]$ for $95\%$ exclusion  & 0.11  & 0.26   & 0.51  & 0.85 & 1.5   & 2.5  & 4.2  & 6.9   & 12  \\
    $Z$ for a $3\,\text{ab}^{-1}$ LHC                  & 10    & 6.7    & 4.8   & 3.7  & 2.8   & 2.2  & 1.7  & 1.3   & 1.0  \\
    \hline
    $\text{BR}({H^{\pm\pm}\rightarrow W^\pm W^\pm})$                                     & 0.44  & 0.55   & 0.64  & 0.73 & 0.84  & 0.96 & 1.1  & 1.2   & 1.4  \\
    signal modifier $\mu$                                           & 0.12  & 0.21   & 0.33  & 0.44 & 0.63  & 0.89 & 1.3  & 1.8   & 2.7  \\
    \hline
  \end{tabular}
  \caption{Signal cross section $\sigma_\text{S}$ and background cross
    section $\sigma_\text{B}$ after cut on $m_T$ (for more details see
    text) for different Higgs mass parameters $m_{H^{\pm\pm}}$. Last
    rows: target luminosity for an $95\%$ exclusion limit, discovery
    significance $Z$ and upper bound on the branching ratio
    $\text{BR}({H^{\pm\pm}\rightarrow W^\pm W^\pm})$ as well as signal
    modifier $\mu$ for a $3\,\text{ab}^{-1}$ LHC.}
  \label{tab:mt_excl}
\end{table}
In addition we quote the discovery significance $Z$ for a
$3\,\text{ab}^{-1}$ LHC, when assuming branching ratios of unity.

\subsection{Branching Ratio $H^{\pm\pm}\rightarrow W^{\pm}W^{\pm}$ and Signal Modifier $\mu$}

We can re-interpret an exclusion of unity branching ratio as a 95\%
confidence level constraint on the branching ratio $H^{\pm\pm}\to
W^\pm W^\pm$. using the cross section for signal and background after
the cuts detailed in the previous section. Eq.~\eqref{eq:target_lumi}
can then be used to arrive at
\begin{align}
  \text{BR}^4_{H^{\pm\pm}\rightarrow W^\pm W^\pm} &= \frac{ 4 \, \left(
    \sigma_\text{S} + \sigma_\text{B} \right) }{ \mathcal{L}
    \sigma_\text{S}^2 }
  \label{eq:gamma}
\end{align}
as the bound that the LHC is sensitive to at a given luminosity
$\mathcal{L}$. Assuming $3~\text{ab}^{-1}$ we present our results in
Tab.~\ref{tab:mt_excl}.
We find that, especially for the low mass regime, branching ratios
significantly smaller than unity can be probed. For the high mass
regime, where the LHC has only little sensitivity, we find branching ratios
greater then unity, signalising that no constraint on the underlying UV structure as motivated in Sec.~\ref{sec:model} can be obtained. 
There a more intuitive expression is the so called
signal modifier $\mu$. Again starting from Eq.~\eqref{eq:target_lumi}
we conclude that
\begin{align}
  \mu &= \frac{Z^2}{2\,\mathcal{L}\,\sigma_\text{S}} + \sqrt{
    \frac{Z^4}{4\,\mathcal{L}^2\,\sigma_\text{S}^2} +
    \frac{Z^2\,\sigma_\text{B}}{\mathcal{L}\,\sigma_\text{S}^2}}\,.
  \label{eq:modif}
\end{align}
Using $Z\equiv 2$ and $\mathcal{L}=3\,\text{ab}^{-1}$ we compute the
upper bound for the signal modifier instead of the branching
ratio. Our results can be found in Tab.~\ref{tab:mt_excl}, too.

\subsection{Statement of Optimality}

When constructing a cut and count analysis it might happen that useful
correlations are not considered. To compute the information which is
accessible when including correlations, we compute the discovery
significance $S/\sqrt{S+B}$ for a $3\,\text{ab}^{-1}$ LHC using the
Fisher discriminant as implemented in
\tmva~\cite{Hocker:2007ht,Speckmayer:2010zz}. In short, the Fisher
discriminant uses the mean value of signal and background as computed
on the domain of observables fed in. A hyperplane sliding along the
line defined by these two points represents a well defined
cut. Therefore, the Fisher discriminant is sensitive to linear
correlations. However, if the means coincide it has no sensitivity at
all even if the shapes of signal and background differ
dramatically. Using a boosted decision tree (BDT), which is also
implemented in \tmva, instead helps identifying such cases. In a
nutshell, a BDT covers the phase space, as parametrised by the
observables we define, with cuboid baskets. These baskets are assigned
either to be background or to be signal. This way irregular shaped
backgrounds or signals can be taken into account. Note, that usually
one cannot easily train a BDT or Fisher discriminant from MC and use
it in an analysis straightforwardly; the propagation of systematic
uncertainties must be taken into account. When there is no connection
to data, as is the case for this purely MC study, this is hard to do
and we stress that we use this tool rather to judge the potential
reach of the observables we have chosen to study. In addition, one
should add that there might exist observables with better
discriminating power than the ones we have chosen so far. The best
discriminating power in identifying these is by comparing sensitivity
yields to the matrix element method~\cite{Kondo:1988yd,Abazov:2004cs},
which is by construction the most discriminating observable between
two specified signal and background hypotheses. Available tools like
\madmax~\cite{Plehn:2013paa}, {\tt{MadWeight}}~\cite{Artoisenet:2010cn},
shower and event deconstruction \cite{Soper:2011cr,Soper:2014rya} have
demonstrated their potential in phenomenological
analyses~\cite{Soper:2012pb,Aad:2016pux,Englert:2015dlp,Khachatryan:2015tzo,Kling:2016lay,FerreiradeLima:2016gcz} and extensions to higher-order matrix elements have been proposed \cite{Campbell:2012cz,Martini:2015fsa,Gritsan:2016hjl}.
The set of observables we use for a proof-of-principle investigation
of the sensitivity reach that can be obtained with these methods are
the ones defined in the previous section. The event selection is
given by Eq.~\eqref{eq:fiducial}. Our results as computed by \tmva~are
presented in Tab.~\ref{tab:tmva}.
\begin{table}[!t]
  \renewcommand{\arraystretch}{1.5}
  \centering
  \begin{tabular}{l|ccccccccc}
    $m_{H}~[\text{GeV}]$    & 200   & 300    & 400   & 500  & 600   & 700  & 800  & 900  & 1000   \\
    \hline
    \hline
    cut and count           & 10    & 6.7    & 4.8   & 3.7  & 2.8   & 2.2  & 1.7  & 1.3  & 1.0  \\
    Fisher discriminant     & 12    & 8.2    & 5.7   & 4.3  & 3.3   & 2.5  & 1.9  & 1.5  & 1.2  \\
    BDT                     & 13    & 8.8    & 6.4   & 4.8  & 3.6   & 2.8  & 2.2  & 1.7  & 1.3  \\
    \hline
  \end{tabular}
  \caption{Discovery significance $S/\sqrt{S+B}$ at a
    $3\,\text{ab}^{-1}$ LHC of our cut and count analysis
    compared to a Fisher discriminant and a BDT for different Higgs
    mass parameters $m_{H^{\pm\pm}}$. Both multivariate analyses use the same
    observables as presented in our cut and count analysis.}
  \label{tab:tmva}
\end{table}
For the Fisher discriminant as well as the BDT we observe a constant
enhancement as function of the Higgs mass compared to our cut and
count analysis of the previous section. Translated into branching
ratios this means that with the help of a BDT, for example, we could
reach $10\%$ better upper bounds than stated in
Tab.~\ref{tab:mt_excl}. In addition the discovery reach for the LHC
cannot be pushed much further than $m_{H^{\pm\pm}}=800$~GeV, even when
accepting the results of a BDT. There is one exception, namely the
$300-400$ GeV mass region. The numbers quoted in Tab.~\ref{tab:tmva}
indicate $20\%$ better bounds on $\text{BR}({H^{\pm\pm}\rightarrow
  W^\pm W^\pm})$ then Tab.~\ref{tab:mt_excl}.

We stress that the dominating background here is of reducible
nature. Indeed, it is dominated by QCD radiation and can therefore be
determined relying on data-driven methods. Due to this fact it might be possible to
train the Fisher discriminant or a BDT on
data~\cite{Bernaciak:2014pna}. In that case the approach would be
robust concerning systematic uncertainties.


\section{Conclusion}
\label{sec:conc}
In this paper we present a detailed study of doubly charged Higgs
production in the limit where the underlying complex triplet has no
relation with electroweak symmetry breaking. In particular, observing
such a particle in its decay to same sign $W$ bosons will provide
important information that might clarify a potential composite nature
of the TeV scale.

We have focussed on 13 TeV LHC collisions and extrapolate to high
luminosity to gauge the extent to which such states can be observed at
the LHC. We have used various analysis strategies to give a more
fine-grained picture of the sensitivity that can be reached in the
light of a small expected inclusive signal. Assuming
$\text{BR}({H^{\pm\pm}\rightarrow W^\pm W^\pm}) \equiv 1.0$ we find
that the LHC is sensitive to heavy Higgs masses up to almost the TeV
regime. A high luminosity LHC of $3\,\text{ab}^{-1}$ is able to
constrain this decay channel for branching ratios of $\sim 0.44$ for
$m_{H^{\pm\pm}} = 200$~GeV growing up to $\sim 0.84$ for
$m_{H^{\pm\pm}} = 600$~GeV using a typical cut and count analysis. We
have demonstrated that using a BDT much better results might be
possible, but only for the mass regime between
$300-400$~GeV. Considering the reducible nature of the dominating
background we encourage experiments to explore this possibility with a
data driven strategy.

A limiting factor of this rather clean, yet rare final state is the
influence of electron charge flips in the high $p_T$ regime and our
scenario provides a motivation to study these effects in same sign,
same flavour leptons on the $Z$ mass pole in association with a high
$p_T$ jet.

\acknowledgments

We would like to thank the MITP for the hospitality and support during the time when
this project was initiated. We would like to thank Luigi Del Debbio, Gabriele Ferretti and Roman Zwicky for helpful conversations
related to the model of Ref.~\cite{Ferretti:2014qta}, and Stefan Richter for
useful discussion on the charge flip of electrons. In particular, we thank Gabriele Ferretti for comments on an earlier version of the manuscript. The work of
P.S. was supported in part by the European Union as part of the FP7
Marie Curie Initial Training Network MCnetITN (PITN-GA-2012-315877).

\bibliography{references}
\bibliographystyle{apsrev4-1}

\end{document}